\documentclass[preprint,12pt,3p]{elsarticle}
\usepackage[utf8x]{inputenc}

\usepackage{amsmath}
\usepackage{amsfonts}
\usepackage{amssymb}

\usepackage{graphicx}

\usepackage[perpage]{footmisc}

\usepackage[bookmarks=true,colorlinks=true,citecolor=blue,linkcolor=red,urlcolor=magenta]{hyperref}

\biboptions{compress}

\usepackage{multirow}

\usepackage[table]{xcolor}

\journal{Annals of Nuclear Energy}

\begin{document}

\begin{frontmatter}

\title{On some fundamental peculiarities of the traveling wave reactor}

\author[ONPU]{V.D.~Rusov\corref{corRusov}}
\cortext[corRusov]{Corresponding author: Vitaliy D. Rusov, E-mail: siiis@te.net.ua}
\author[ONPU]{V.A.~Tarasov}
\author[ONPU]{I.V.~Sharph}
\author[SEA]{V.N.~Vaschenko}
\author[ONPU]{E.P.~Linnik}
\author[ONPU]{T.N.~Zelentsova}
\author[ONPU]{R.~Beglaryan}
\author[ONPU]{S.~Chernegenko}
\author[ONPU]{S.I.~Kosenko}
\author[ONPU]{V.P.~Smolyar}

\address[ONPU]{Department of Theoretical and Experimental Nuclear Physics, \\ Odessa  National Polytechnic University, Odessa, Ukraine}
\address[SEA]{State Ecological Academy for Postgraduate Education, Kiev, Ukraine}

\begin{abstract}
On the basis of the condition for nuclear burning wave existence in the 
neutron-multiplicating media (U-Pu and Th-U cycles) we show the possibility of 
surmounting the so-called dpa-parameter problem, and suggest an algorithm of the
optimal nuclear burning wave mode adjustment, which is supposed to yield the 
wave parameters (fluence/neutron flux, width and speed of nuclear burning wave) 
that satisfy the dpa-condition associated with the tolerable level of the 
reactor materials radioactive stability, in particular that of the cladding 
materials.

It is shown for the first time that the capture and fission cross-sections of 
$^{238}$U and $^{239}$Pu increase with temperature within 1000-3000~K range, 
which under certain conditions may lead to a global loss of the nuclear burning
wave stability. Some variants of the possible stability loss due to the 
so-called blow-up modes (anomalous nuclear fuel temperature and neutron flow 
evolution) are discussed and are found to possibly become a reason for a trivial
violation of the traveling wave reactor internal safety.
\end{abstract}


\end{frontmatter}

\section{Introduction}
\label{sec1}

Despite the obvious and unique effectiveness of nuclear energy of the new 
generation, there are difficulties of its understanding related to the 
nontrivial properties of an ideal nuclear reactor of the future.

First, nuclear fuel should be natural, i.e. non-enriched uranium or thorium. 
Second, traditional control rods should be absolutely absent in reactor active 
zone control system. Third, despite the absence of the control rods, the 
reactor must exhibit the so-called internal safety. This means that under any 
circumstances the reactor active zone must stay at a critical state, i.e. 
sustain a normal operation mode automatically, with no operator actions, 
through physical causes and laws, that naturally prevent the explosion-type 
chain reaction. Figuratively speaking, the reactors with internal safety are
"the nuclear devices that never explode" \cite{ref1}.

Surprisingly, reactors that meet such unusual requirements are really possible. 
The idea of such self-regulating fast reactor was expressed for the first time
in a general form (the so-called breed-and-burn mode) by Russian physicists 
Feynberg and Kunegin during the II Geneva conference in 1958~\cite{ref2} and 
was relatively recently "reanimated" in a form of the self-regulating fast 
reactor in traveling nuclear burning wave mode by Russian physicist 
Feoktistov~\cite{ref3} and independently by American physicists Teller, 
Ishikawa and Wood~\cite{ref4}.

The main idea of the reactor with internal safety is that the fuel components 
are chosen in such a way that, first, the characteristic time $\tau_{\beta}$ of 
the active fuel component (the fissile component) nuclear burning is
significantly larger than the time of the delayed neutrons appearance; and 
second, all the self-regulation conditions are sustained in the operation mode. 
Particularly, the equilibrium concentration $\tilde{n}_{fis}$ of the active fuel
component, according to Feoktistov's condition of the wave mode existence, is 
greater than its critical concentration\footnote{Concentrations of the active 
element ($^{239}Pu$ and $^{233}U$ in cycles (\ref{eq1a}) and (\ref{eq2a})), are 
called equilibrium or critical when an equal number of the active element nuclei
or neutrons, respectively, is born and destroyed at the same time during the 
nuclear cycle.} $n_{crit}$ \cite{ref3}. These conditions are very important, 
though they are almost always practically implementable in case when the 
nuclear transformations chain of Feoktistov's uranium-plutonium cycle 
type~\cite{ref3} is significant among other reactions in the reactor:

\begin{equation}
  {^{238} U (n,\gamma)} \rightarrow {^{239} U} \xrightarrow[]{\beta^{-}} {^{239} Np} \xrightarrow[]{\beta^{-}} {^{239} Pu (n,fission)}
\label{eq1a}
\end{equation}

The same is also true for  the Teller-Ishikawa-Wood thorium-uranium cycle 
type~\cite{ref4}

\begin{equation}
{  ^{232} Th (n,\gamma)} \rightarrow {^{233} Th} \xrightarrow[]{\beta^{-}} {^{233} Pa} \xrightarrow[]{\beta^{-}} {^{233} U (n,fission)},
\label{eq2a}
\end{equation}

In these cases the fissionable isotopes form ($^{239}Pu$ in~(\ref{eq1a}) or 
$^{233}U$ in~(\ref{eq2a})) which are the active components of the nuclear fuel. 
The characteristic time of such reaction depends on the time of the 
corresponding $\beta$-decays, and approximately equals to 
$\tau_{\beta} = 2.3 / ln{2} \approx 3.3$ days in case~(\ref{eq1a}) and 
$\tau_{\beta} \approx 39.5$ days in case~(\ref{eq2a}) which is many orders of 
magnitude higher than the corresponding time for the delayed neutrons.

The effect of the nuclear burning process self-regulation is provided by the 
fact that the system, being left by itself, cannot surpass the critical state 
and enter the uncontrolled reactor runaway mode, because the critical 
concentration is limited from above by a finite value of the active fuel 
component equilibrium concentration (plutonium in~(\ref{eq1a}) or uranium 
in~(\ref{eq2a})): $\tilde{n}_{fis} > n_{crit}$ (the Feoktistov's wave existence condition~\cite{ref3}).

Phenomenologically the process of the nuclear burning self-regulation is as 
follows. Any increase in neutron flow leads to a quick burn-out of the active 
fuel component (plutonium in~(\ref{eq1a}) or uranium in~(\ref{eq2a})), i.e. 
to a reduction of their concentration and neutron flow; meanwhile the formation 
of the new nuclei by the corresponding active fuel component proceeds with the 
prior rate during the time $\tau_{\beta}$. On the other hand, if the neutron 
flow drops due to some external impact, the burn-out speed reduces and the 
active component nuclei generation rate increases, followed by the increase of 
a number of neutrons generated in the reactor during approximately the same time
$\tau_{\beta}$.

\begin{figure}
  \begin{center}
    \includegraphics[width=10cm]{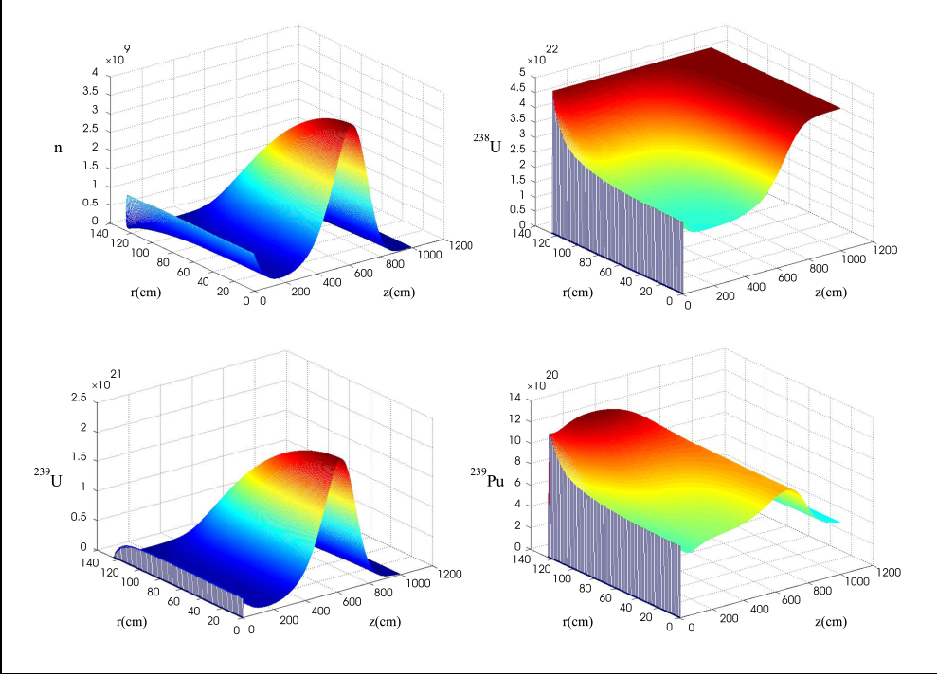}
    \caption{Kinetics of the neutrons, $^{238}$U, $^{239}$U and $^{239}$Pu 
    concentrations in the core of a cylindrical reactor with radius of 125~cm 
    and 1000~cm long at the time of 240~days. Here $r$ is the transverse 
    spatial coordinate axis (cylinder radius), $z$ is the longitudinal spatial 
    coordinate axis (cylinder length). Temporal step of the numerical 
    calculations is 0.1~s. Adopted from \cite{ref29}}
    \label{fig1}
  \end{center}
\end{figure}

The system of kinetic equations for nuclei (the components of nuclear fuel) and 
neutrons (in diffuse approximation) in such chains are rather simple. They 
differ only by the depth of description \cite{ref3,ref4,ref5,ref6,ref7,ref8,
ref9,ref10,ref11,ref12,ref13,ref14,ref15,ref16,ref17,ref18,ref18a,ref19,ref20,
ref21,ref22,ref23,ref24,ref25,ref26,ref27,ref28,ref29} of all the possible 
active fuel components and non-burnable poison\footnote{Here by poison we mean 
the oxygen nuclei or other elements, chemically bound to heavy nuclides, 
construction materials, coolant and the poison itself, i.e. the nuclei added to 
the initial reactor composition in order to control the neutron balance.}. 
Fig.~\ref{fig1} shows the characteristic solutions for such problem (equations 
(3)-(9) in \cite{ref29}) in a form of the soliton-like waves of the nuclear 
fuel components and neutrons concentrations for uranium-plutonium cycle in a
cylindrical geometry case. Within the theory of the soliton-like fast reactors 
it is easy to show that in general case the phase speed $u$ of soliton-like 
neutron wave of nuclear burning is defined by the following approximate 
equality \cite{ref29}:

\begin{equation}
  \frac{u \tau_{\beta}}{2L} \simeq \left( \frac{8}{3 \sqrt{\pi}} \right)^6 \exp {\left( - \frac{64}{9 \pi} a^2 \right)}, ~~ a^2 = \frac{\pi ^2}{4} \cdot \frac{n_{crit}}{\tilde{n}_{fis} - n_{crit}},
\label{eq3a}
\end{equation}

\noindent where $\tilde{n}_{fis}$ and $n_{crit}$ are the equilibrium and 
critical concentration of the active (fissile) isotope, $L$ is the mean neutron 
diffusion length, $\tau_{\beta}$ is the delay time, associated with the birth 
of the active (fissile) isotope and equal to an effective $\beta$-decay period 
of the intermediate nuclei in Feoktistov's uranium-plutonium cycle (\ref{eq1a}) 
or in Teller-Ishikawa-Wood thorium-uranium cycle (\ref{eq2a}).

Let us note that expression (\ref{eq3a}) automatically incorporates a condition 
of nuclear burning process self-regulation, since the fact of a wave existence 
is obviously predetermined by the inequality $\tilde{n}_{fis} > n_{crit}$. In 
other words, the expression (\ref{eq3a}) is a necessary physical condition of 
the soliton-like neutron wave existence. Let us note for comparison that the 
maximal value of the nuclear burning wave phase velocity, as follows from 
(\ref{eq3a}), is characterized both for uranium and thorium by the equal 
average diffusion length ($L \sim 5 cm$) of the fast neutrons ($1 ~MeV$) and 
is equal to $3.70 ~cm / day$ for uranium-plutonium cycle (\ref{eq1}) and 
$0.31~cm/day$ for thorium-uranium cycle (\ref{eq2a}).

Generalizing the results of a wide range of numerical experiments\cite{ref5,
ref6,ref7,ref8,ref9,ref12,ref14,ref16,ref18,ref18a,ref19,ref20,ref21,ref22,
ref23,ref26,ref27,ref28,ref29}, we can positively affirm that the principal
possibility of the main stationary wave parameters control was reliably 
established within the theory of a self-regulating fast reactor in traveling 
wave mode, or in other words, the traveling wave reactor (TWR). It is possible 
both to increase the speed, the thermal power and the final fluence as well as 
decrease them. Obviously, according to~(\ref{eq3a}), it is achieved by varying 
the equilibrium and critical concentrations of the active fuel component, i.e. 
by the purposeful change of the initial nuclear fuel composition.

The technological problems of TWR are actively discussed in science nowadays. 
The essence of these problems usually comes to a principal impossibility of such
project realization, and is defined by the following insurmountable flaws: 
\begin{itemize}
\item High degree of nuclear fuel burn-up (over 20\% in average) leading to the 
      following adverse consequences:
\begin{itemize}
    \item High damaging dose of fast neutrons acting at at the constructional 
          materials ($\sim$500~$dpa$)\footnote{For comparison -- the claimed
	  parameters for the Russian FN-800 reactor are $93$~$dpa$. At the same
	  time it is known that one of the main tasks of the construction 
	  materials radioactive stability investigations conducted at the 
	  Bochvar Hi-tech Institute for non-organic materials (Moscow) is 
	  to achieve $133$ to $164$~$dpa$ by 2020!};
    \item High gas release, which requires an increased gas cavity length on top
          of a long fuel rod as it is.
\end{itemize}
\item Long active zone requiring the correspondingly long fuel rods, which makes
their parameters unacceptable from the technological use point of view. For 
instance, this has to do with the parameters characterized by a significant 
increase in:
\begin{itemize}
\item the value of a positive void coefficient of reactivity;
\item hydraulic resistance;
\item energy consumption for the coolant circulation through the reactor.
\end{itemize}
\item The problem of nuclear waste associated with the unburned plutonium 
      reprocessing and nuclear waste utilization.
\end{itemize}

The main goal of the present paper is to solve the specified technological 
problems of the TWR on the basis of a technical concept which makes it 
impossible for the damaging dose of the fast neutrons in the reactor (fuel rods 
jackets, reflection shield and reactor pit) to exceed the $\sim 200~dpa$ level. 
The essence of this technical concept is to provide the given neutron flux on 
in-reactor devices by defining the speed of the fuel movement relative to the
nuclear burning wave speed. The neutron flux, wave speed and fuel movement speed
are in their turn predetermined by the chosen parameters (equilibrium and 
critical concentrations of the active component in the initial nuclear fuel 
composition).

Section~\ref{sec1} of this paper is dedicated to a brief analysis of the 
state-of-the-art idea of a self-regulating fast reactor in traveling wave mode.
Based on this analysis we formulate the problem statement and chalk out the 
possible ways to solve it. Chapter~\ref{sec2} considers the analytical solution 
for a non-stationary 1D reactor equation in one-group approximation with 
negative reactivity feedback (1D Van~Dam~\cite{ref7} model). It yields the 
expressions for the amplitude $\varphi_m$, phase $\alpha$ and phase speed $u$, 
as well as the dispersion (FWHM) of the soliton-like burning wave. Knowing the 
FWHM we may further estimate the spatial distribution of the neutron flux and 
thus a final neutron fluence. Chapter~\ref{sec3} is dedicated to a description
of the nontrivial neutron fluence dependence on phase velocity of the solitary 
burn-up waves in case of the fissible and non-fissible absorbents. It reveals 
a possibility of the purposeful (in terms of the required neutron fluence and 
nuclear burning wave speed values) variation of the initial nuclear fuel 
composition. Chapter~\ref{sec4} analyses the dependence of the damaging dose on 
neutron fluence, phase velocity and dispersion of the solitary burn-up waves.
Chapter~\ref{sec5} considers the possible causes of the TWR internal safety 
violation caused by ``Fukushima plutonium'' effect, or in other words, the 
temperature blow-up modes driven either by temperature or neutron flux. 
Chapter~\ref{sec6} is dedicated to analysis of the practical examples of the 
temperature blow-up modes in neutron-multiplying media. The idea of an impulse 
thermo-nuclear TWR is also proposed. The conclusion of the paper is presented 
in Chapter~\ref{sec7}.

\section{On entropy and dispersion of solitary burn-up waves.}
\label{sec2}

Let us discuss the physical causes, defining the main characteristics of 
the soliton-like propagation of ``criticality'' wave in the initially 
undercritical environment, characterized by the infinite multiplication factor 
$k_{\infty}$ less than unit. Obviously, the supercritical area  
($k_{\infty} > 1$) must be created by some external neutron source (e.g. by an 
accelerator or another super-critical area). In the general case, the 
supercritical area is a result of the breeding effects in fast nuclear systems 
or the burning of the fissible absorbents (fuel components) in thermal nuclear 
systems\footnote{W.~Seifritz (1995) was the first to find theoretically a 
nuclear solitary burn-up wave in opaque neutron absorbers~\cite{ref30}. The 
supercriticality waves in thermal nuclear reactors are searched for and analysed
in the papers by Akhiezer~A.I. et al.\cite{ref31,ref32,ref33,ref34}, where they
show the possibility of both fast \cite{ref31,ref32,ref33} and slow~\cite{ref34}
modes of nuclear burning distribution (i.e. the super-criticality waves) in the
framework of diffuse approximation.}. Due to the gradual burn-out of the 
neutron-multiplying medium in the supercritical area, this area looses its 
supercritical properties, since $k_{\infty}$ becomes less than unit. The wave
would have to stop and diminish at this point in an ordinary case. However, 
because of the neutrons, appearing during breeding and diffusely "infecting" the
nearby areas, this "virgin" area before the wavefront is forced to obtain the 
properties of super-criticality, and the wave moves forward in this direction. 
Apparently, the stable movement of such soliton-like wave requires some kind of 
stabilizing mechanism. For example, the negative self-catalysis or any other 
negative feedback. This is called the negative reactivity feedback in 
traditional nuclear reactors\footnote{According to Van~Dam~\cite{ref7}, the 
procedure of the reactivity introduction into the 1D non-stationary equation of
the reactor in one-group approximation, though implicitly, takes the kinetic 
equations of the burn-out into account. Particularly, the production of 
plutonium in U-Pu cycle or uranium in Th-U cycle.}. Let us therefore consider
such an example qualitatively below.

For this purpose let us write down a 1D non-stationary equation of the reactor 
in one-group approximation \cite{ref35,ref36,ref37} with negative reactivity 
feedback:

\begin{equation}
  D \frac{\partial ^2 \varphi}{\partial x^2} + \left( k_{\infty} -1 + \gamma \varphi \right) \Sigma_{a} \varphi = \frac{1}{v} \frac{\partial \varphi}{\partial t}
\label{eq1}
\end{equation}

\noindent where $\varphi$ is the neutron flux $[cm^{-2} s^{-1}]$; $D$ is the 
diffusion coefficient, $[cm]$; $\gamma \varphi$ is the reactivity, 
dimensionless value; $\Sigma_a$ is the total macroscopic absorption 
cross-section, $[cm^{-1}]$; $v$ is the neutron speed, $[cm \cdot s^{-1}]$. In 
this case the negative feedback $\gamma$ is defined mainly by the fact that the 
infinite multiplication factor is used in (\ref{eq1}), and therefore the flux 
density must be corrected.

Let us search for the solution in an autowave form:

\begin{equation}
\varphi (x,t) = \varphi (x - ut) \equiv \varphi (\xi),
\label{eq2}
\end{equation}

\noindent where $u$ is the wave phase velocity, $\xi$ is the coordinate in a 
coordinate system, which moves with phase speed. In such case:

\begin{equation}
\frac{1}{v} \frac{\partial \varphi}{\partial t} = - \frac{u}{v} \frac{\partial \varphi}{\partial t}
\label{eq3}
\end{equation}

As is known \cite{ref7}, the relation $u/v$ by order of magnitude equals to 
$10^{-13}$ and $10^{-11}$ for fast and thermal nuclear systems respectively. 
Therefore the partial time derivative in (\ref{eq1}) may be neglected without 
loss of generality. Further taking into account (\ref{eq2}), the equation 
(\ref{eq1}) may be presented in the following form:

\begin{equation}
 L^2 \frac{\partial ^2 \varphi}{\partial \xi ^2} + \left[ k_{\infty} (\psi) - 1 + \gamma \varphi \right] \varphi = 0,
\label{eq4}
\end{equation}

\noindent where $L = (D / \Sigma_a) ^{1/2}$ is the neutron diffusion length, 
and $\psi$ is the so-called neutron fluence function:

\begin{equation}
\psi (x,t) = \int \limits _0 ^t \varphi (x,t') dt'.
\label{eq5}
\end{equation}

In order to find a physically sensible analytic solution of (\ref{eq4}) by 
substituting (\ref{eq5}), we need to define some realistic form of the function 
$k_{\infty} (\psi)$ (usually referred to as the burn-up function). Since the 
real burn-up function $k_{\infty} (\psi)$ has a form of some asymmetric 
bell-shaped dependence on fluence, normalized to its maximal value $\psi_{max}$
(fig.~\ref{fig1}), following \cite{ref7}, let us define it in a form of 
parabolic dependency without lose of generality:

\begin{equation}
  k_{\infty} = k_{max} + \left( k_0  - k_{max} \right) \left( \frac{\psi}{\psi_{max}} -1 \right)^2,
\label{eq6}
\end{equation}

\noindent where $k_{max}$ and $k_0$ are the maximal and initial neutron 
multiplication factors. Substituting (\ref{eq6}) into (\ref{eq4}) we obtain:

\begin{equation}
L^2 \varphi_{\xi \xi} + \rho_{max} \varphi + \gamma_0 \varphi^2 - \delta \left[ \frac{\int \limits_\xi ^{\infty} \varphi d\xi}{u \psi_m} - 1 \right]^2 \varphi
 = 0,
\label{eq7}
\end{equation}

\noindent where $\rho_{max} = k_{max} -1$, $\delta = k_{max} - k_0$, $\gamma_0 \equiv \gamma$.

\begin{figure}
  \begin{center}
    \includegraphics[width=10cm]{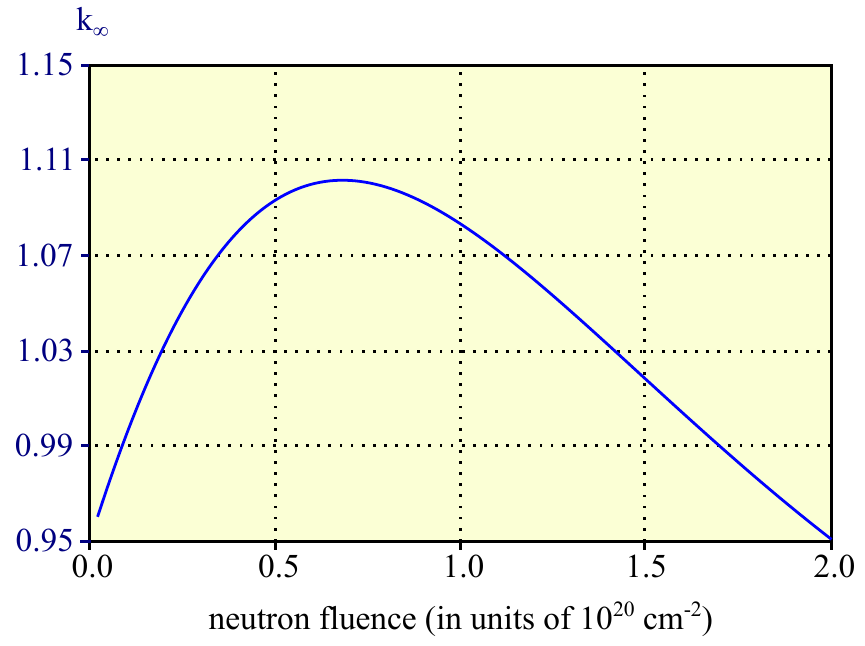}
    \caption{Asymmetric burn-up function as characteristic for realistic burn-up function. Adapted from \cite{ref7}.}
    \label{fig2}
  \end{center}
\end{figure}

Suppose we are searching a partial solution of (\ref{eq7}). Let us rewrite it
in the following form:

\begin{equation}
  L^2 \frac{d^2 \varphi}{d \xi^2} + \left( \rho_{max} + \gamma_0 \varphi - \delta \left( \frac{\int \limits _{\xi} ^{+\infty} \varphi d \xi}{u \psi_m} - 1  \right)^2 \right) \varphi = 0.
\label{eq8}
\end{equation}

Let us introduce a new unknown function:

\begin{equation}
\chi (\xi) = \int \limits _{\xi} ^{+\infty} \varphi d \xi \Rightarrow \varphi (\xi) = - \frac{d \chi (\xi)}{d \xi},
\label{eq9}
\end{equation}

\noindent that due to its non-negativity on the interval $\xi \in [0,\infty]$, 
must satisfy the following boundary conditions: $\varphi = 0$ for 
$\xi = 0,\infty$.

The equation will take the form:

\begin{equation}
 L^2 \frac{d^3 \chi (\xi)}{d \xi ^3} + \left( \rho_{max} - \gamma_0 \frac{d \chi (\xi)}{d \xi} - \delta \left( \frac{\chi (\xi)}{u \psi_m} -1 \right)^2 \right) \frac{d \chi (\xi)}{d \xi} = 0.
\label{eq10}
\end{equation}

In order to find a partial solution of (\ref{eq10}), we require the following 
additional condition to hold:

\begin{equation}
\rho_{max} - \gamma_0 \frac{d \chi (\xi)}{d \xi} - \delta \left( \frac{\chi (\xi)}{u \psi_m} - 1 \right)^2 = f \left( \chi (\xi) \right),
\label{eq11}
\end{equation}

\noindent where $f (\xi)$ is an arbitrary function, the exact form of which will
be defined later. The condition (\ref{eq11}) is chosen because it makes it 
possible to integrate the equation (\ref{eq10}). Really, if (\ref{eq11}) is 
true, the equation (\ref{eq2}) takes the following form:

\begin{equation}
  L^2 \frac{d^3 \chi (\xi)}{d \xi^3} + f \left( \chi (\xi) \right) \frac{d \chi (\xi)}{d \xi} = 0,
\label{eq12}
\end{equation}

That allows us reduce the order of the equation:

\begin{equation}
L^2 \frac{d^2 \chi (\xi)}{d \xi ^2} + F_1 \left( \chi (\xi) \right) = C.
\label{eq13}
\end{equation}

Here $F_1 (\chi)$ denotes a primitive of $f(\chi)$, and $C$ is an arbitrary 
integration constant. The order of (\ref{eq13}) may be further reduced by 
multiplying both sides of the equation by $d \chi (\xi) / d \xi$:

\begin{equation}
\frac{L^2}{2} \left( \frac{d \chi (\xi)}{d \xi} \right)^2 + F_2 \left( \chi (\xi) \right) - C \chi (\xi) = B,
\label{eq14}
\end{equation}

$F_2(\chi)$ here denotes a primitive of $F_1(\chi)$, i.e. ``the second 
primitive'' of the function $f(\chi)$ introduced in (\ref{eq2}), and $B$ is a 
new integration constant.

The obtained equation (\ref{eq14}) is a separable equation and may be rewritten 
in the following form:

\begin{equation}
  d \xi = \pm \frac{d \chi}{\sqrt{\frac{2}{L^2} \left( B - F_2 \left( \chi (\xi) \right) + C \chi (\xi) \right)}}.
\label{eq15}
\end{equation}

On the other hand, (\ref{eq11}) may also be considered a separable equation 
relative to $\chi (\xi)$. Then, separating variables in (\ref{eq11}) we obtain:

\begin{equation}
d \xi = \frac{d \chi}{\frac{\rho_{max}}{\gamma_0} - \frac{\delta}{\gamma_0} \left( \frac{\chi (\xi)}{u \psi_m} - 1 \right)^2 - \frac{1}{\gamma_0} f \left( \chi (\xi) \right)}.
\label{eq16}
\end{equation}

Since the equations (\ref{eq15}) and (\ref{eq16}) are for the same function 
$\chi (\xi)$, by comparing them, we derive that the following relation must 
hold:

\begin{equation}
\pm \sqrt{\frac{2}{L^2} \left( B - F_2 \left( \chi (\xi) \right) + C \chi (\xi) \right)} = \frac{\rho_{max}}{\gamma_0} - \frac{\delta}{\gamma_0} \left( \frac{\chi (\xi)}{u \psi_m} -1 \right)^2 - \frac{1}{\gamma_0} f \left( \chi (\xi) \right).
\label{eq17}
\end{equation}

In order to simplify (\ref{eq15}) and (\ref{eq16}), let us choose $f(\chi)$ in a
polynomial form of $\chi$. The order of this polynomial is $n$. Then 
$F_2 (\xi)$, obtained by double integration of $f(\chi)$, is a polynomial of 
order $(n+2)$. Taking the square root, according to (\ref{eq17}), should also 
lead to a polynomial of the order $n$. Therefore $n + 2 = 2n \Rightarrow n = 2$.

Consequently, the function $f(\chi)$ may only be a second-order polynomial under
the assumptions made above.

\begin{equation}
f (\chi) = s_2 \chi^2 + s_1 \chi + s_0,
\label{eq18}
\end{equation}

\noindent where $s_0$, $s_1$, $s_2$ are the polynomial coefficients.

Double integration of (\ref{eq18}) leads to:

\begin{equation}
F_2 (\chi) = \frac{s_2}{12} \chi^4 + \frac{s_1}{6} \chi^3 + \frac{s_0}{2} \chi^2 +c_1 \chi + c2,
\label{eq19}
\end{equation}

\noindent where $c_1$ and $c_2$ are the integration constants. 

Substituting (\ref{eq18}) and (\ref{eq19}) into (\ref{eq17}) we get:

\begin{align}
& \left(\frac{\rho_{max}}{\gamma_0} - \frac{\delta}{\gamma_0} \left( \frac{\chi}{u \psi_m} - 1 \right)^2 - \frac{1}{\gamma_0} \left( s_2 \chi^2 + s_1 \chi + s_0 \right) \right)^2 = \nonumber \\
& = \frac{2}{L^2} \left( B - \left( \frac{s_2}{12} \chi^4 + \frac{s_1}{6} \chi^3 + \frac{s_0}{2} \chi^2 + c_1 \chi + c_2 \right) + C \chi \right)
\label{eq20}
\end{align}

Further in (\ref{eq20}) we set the coefficients at the same orders of $\chi$ 
equal:

\begin{equation}
\left( \frac{\delta}{\gamma_0 u^2 \psi_m^2} + \frac{s_2}{\gamma_0} \right)^2 = -\frac{s_2}{6L^2},
\label{eq21}
\end{equation}

\begin{equation}
-2 \left( \frac{\delta}{\gamma_0 u^2 \psi_m^2} + \frac{s_2}{\gamma_0} \right) \left( \frac{2\delta}{\gamma_0 u \psi_m} - \frac{s_1}{\gamma_0} \right) = - \frac{s_1}{3L^2},
\label{eq22}
\end{equation}

\begin{equation}
\left( \frac{2 \delta}{\gamma_0 u \psi_m} - \frac{s_1}{\gamma_0} \right)^2 - 2 \left( \frac{\delta}{\gamma_0 u^2 \psi_m^2} + \frac{s_2}{\gamma_0} \right) \left( \frac{\rho_{max}}{\gamma_0} - \frac{\delta}{\gamma_0} - \frac{s_0}{\gamma_0} \right)  = - \frac{s_0}{L^2};
\label{eq23}
\end{equation}

\begin{equation}
2 \left( \frac{2\delta}{\gamma_0 u \psi_m} - \frac{s_1}{\gamma_0} \right) \left( \frac{\rho_{max}}{\gamma_0} - \frac{\delta}{\gamma_0} - \frac{s_0}{\gamma_0} \right) = - \left( \frac{2c_1}{L^2} + \frac{2C}{L^2} \right),
\label{eq24}
\end{equation}

\begin{equation}
\left( \frac{\rho_{max}}{\gamma_0} - \frac{\delta}{\gamma_0} - \frac{s_0}{\gamma_0} \right)^2 = \frac{2}{L^2}B - \frac{2c_2}{L^2}.
\label{eq25}
\end{equation}

Note that the first three equations are enough to find the coefficients $s_0$, 
$s_1$ and $s_2$, and the remaining two equations may be satisfied with the 
appropriate constants $B$, $C$, $c_1$, $c_2$.

\begin{equation}
s_0 = \rho_{max} - \delta;
\label{eq26}
\end{equation}

\begin{equation}
s_1 = \frac{2\delta}{u \psi_m} - \frac{\gamma_0}{L} \sqrt{\delta - \rho_{max}};
\label{eq27}
\end{equation}

\begin{equation}
s_2 = \frac{\delta \gamma_0}{3 L u \psi_m \sqrt{\delta - \rho_{max}}} - \frac{\gamma_0 ^2}{6L^2} - \frac{\delta}{u^2 \psi_m ^2}.
\label{eq28}
\end{equation}

After finding $s_0$, $s_1$, $s_2$ from this system, we may consider the equation
(\ref{eq16}) in more detail, which takes the form:

\begin{equation}
d \xi = \frac{d \chi}{- \left( \frac{\delta}{\gamma_0 u^2 \psi_m ^2} + \frac{s_2}{\gamma_0} \right) \chi ^2 + \left( \frac{2 \delta}{\gamma_0 u \psi_m} + \frac{s_1}{\gamma_0} \right) \chi + \left( \frac{\rho_{max}}{\gamma_0} - \frac{\delta}{\gamma_0} - \frac{s_0}{\gamma_0} \right)}.
\label{eq29}
\end{equation}

Solving this equation yields:

\begin{equation}
\frac{d \chi}{\left( \chi - K \right)^2 - M^2} = - N d \xi,
\label{eq30}
\end{equation}

\noindent where

\begin{equation}
K = \frac{\frac{\delta}{\gamma_0 u \psi_m} - \frac{s_1}{2 \gamma_0}}{\frac{\delta}{\gamma_0 u^2 \psi_m^2} + \frac{s_2}{\gamma_0}},
\label{eq31}
\end{equation}

\begin{equation}
M^2 = \left( \frac{\frac{\delta}{\gamma_0 u \psi_m} - \frac{s_1}{2\gamma_0}}{\frac{\delta}{\gamma_0 u^2 \psi_m^2} + \frac{s_2}{\gamma_0}} \right) + \frac{\left(\frac{\rho_{max}}{\gamma_0} - \frac{\delta}{\gamma_0} - \frac{s_0}{\gamma_0} \right)}{\left( \frac{\delta}{\gamma_0 u^2 \psi_m^2} + \frac{s_2}{\gamma_0} \right)},
\label{eq32}
\end{equation}

\begin{equation}
N = \frac{\delta}{\gamma_0 u^2 \psi_m^2} + \frac{s_2}{\gamma_0}.
\label{eq33}
\end{equation}

Let us introduce a new variable $\chi_1$ into (\ref{eq30}) by substituting:

\begin{equation}
\chi - K = M \chi_1, ~~ d \chi = M d\chi_1.
\label{eq34}
\end{equation}

Then the equation (\ref{eq30}) will take the following form:

\begin{equation}
\frac{d \chi_1}{(\chi_1)^2 - 1} = - MN d\xi.
\label{eq35}
\end{equation}

Hence

\begin{equation}
\chi_1 = - \tanh {(MN \xi - D)},
\label{eq36}
\end{equation}

\noindent where $D$ is the integration constant. Taking into account that
$\chi - K = M \chi_1$:

\begin{equation}
\chi = K - M \tanh {(MN \xi - D)}.
\label{eq37}
\end{equation}

Considering (\ref{eq9}), we obtain the soliton-like solution in the form:

\begin{equation}
\varphi (\xi) = M^2 N \sec {h^2 (MN \xi - D)}.
\label{eq38}
\end{equation}

Let us remind that together with introducing a new unknown function $\chi (\xi)$
(see (\ref{eq9})) we obtained an obvious condition for this function:

\begin{equation}
\lim \limits_{\xi \rightarrow \infty} \chi (\xi) = 0.
\label{eq39}
\end{equation}

Let us show that this condition eventually leads to an autowave existence 
condition. Obviously the condition (\ref{eq39}) along with (\ref{eq37}) 

\begin{equation}
\lim \limits_{\xi \rightarrow \infty} \chi (\xi) = \lim \limits_{\xi \rightarrow \infty} \left[ K - M \tanh {(MN \xi - D)} \right] = 0
\label{eq40}
\end{equation}

\noindent leads to

\begin{equation}
K = M.
\label{eq41}
\end{equation}

This relation lets us define the amplitude $\varphi_m$, phase $\alpha$ and 
phase velocity $u$ of the soliton-like wave:

\begin{equation}
\alpha = MN = \frac{\sqrt{\delta - \rho_{max}}}{2L};
\label{eq42}
\end{equation}

\begin{equation}
\varphi_m = M^2 N = \frac{\delta - 3 \rho_{max}}{2\gamma_0} = \frac{3 \rho_{max} - \delta}{2 \vert \gamma_0 \vert};
\label{eq43}
\end{equation}

\begin{equation}
u = \frac{\varphi_m}{\alpha \psi_m}.
\label{eq44}
\end{equation}

From the condition of non-negative width (\ref{eq42}) and amplitude (\ref{eq43})
of the nuclear burning wave, follows the condition of 1D autowave existence, or
the so-called ``ignition condition'' by van~Dam~\cite{ref7}:

\begin{equation}
3 \rho_{max} - \delta = 2k_{max} + k_0 - 3 \geqslant 0, ~~~ where~ 1 - k_0 > 0.
\label{eq45}
\end{equation}

It is noteworthy that the analogous results for a nonlinear one-group diffusion 
1D-model (\ref{eq1}) with explicit feedback and burn-up effects were first 
obtained by Van~Dam\cite{ref7}. The same results (see (\ref{eq42})-(\ref{eq44}))
were obtained by Chen and Maschek~\cite{ref12} while investigating the 3D-model 
by Van~Dam using the perturbation method. The only difference is that the value 
of neutron fluence associated with the maximum of burn-up parameter $k_{\infty}$
was adapted to the transverse buckling. In other words, they considered the 
transverse geometric buckling mode as a basis for perturbation. Hence they 
introduced a geometric multiplication factor $k_{GB}$ due to transverse buckling
into two-dimensional equation (\ref{eq1}), which led to a change in some initial
parameters ($\rho_{max} = k_{max} - k_{GB}, ~\delta = k_{max} - k_0$) and 
consequently -- to a change in the conditions of the autowave existence in 3D 
case:

\begin{equation}
3 \rho_{max} - \delta = 2k_{max} + k_0 - 3k_{GB} \geqslant 0, ~~~ where~ k_{GB} - k_0 > 0,
\label{eq46}
\end{equation}

\noindent that in the case of $k_{GB} = 1$ is exactly the same as the so-called 
ignition condition by Van~Dam~\cite{ref7}.

From the point of view of the more detailed Feoktistov model~\cite{ref3} 
analysis, thoroughly considered in \cite{ref29} and related to a concept of 
the nuclear systems internal safety, the condition (\ref{eq46}) is necessary, 
but not sufficient. On the other hand, it is an implicit form of the necessary 
condition of wave existence according to Feoktistov, where the equilibrium 
concentration $\tilde{n}_{fis}$ of the active fuel component must be greater 
than its critical concentration $n_{crit}$ ($\tilde{n}_{fis} > n_{crit}$)~
\cite{ref3,ref29}. The physics of such hidden but simple relation will be 
explained below (see Chapter~\ref{sec3}).

Returning to a 1D reactor equation solution (\ref{eq7}) in one-group 
approximation with negative reactivity feedback, let us write it in a more 
convenient form for analysis

\begin{equation}
\varphi = \varphi_m \sec {h^2 (\alpha \xi)} = \varphi_m \sec {h^2 \left[ \alpha ( x - ut) \right]}.
\label{eq46a}
\end{equation}

\noindent where $\varphi_m$ is the amplitude of the neutron flux; $1/ \alpha$ is
the characteristic length proportional to the soliton wave width, which is a 
full width at half-maximum (FWHM) by definition, and equals to

\begin{equation}
\Delta _{1/2} = FWHM = 2 \ln{ \left( 1 + \sqrt{2} \right)} \alpha ^{-1}, ~~[cm].
\label{eq47}
\end{equation}

Apparently, integrating (\ref{eq46}) yields the area under such soliton:

\begin{equation}
A_{area} = 2 \varphi_m / \alpha, ~~ [cm^2].
\label{eq48}
\end{equation}

In order to estimate the extent of the found parameters influence on the 
dynamics of the soliton-like nuclear burning wave stability, let us invoke an 
information-probability approach, developed by Seifritz~\cite{ref38}. For this
purpose let us write down the expression for the mean value of information or 
more precisely -- the entropy of the studied process:

\begin{equation}
S = - k_B \int \limits _{-\infty} ^{\infty} p (x) \ln {p(x)} dx,
\label{eq49}
\end{equation}

\noindent where $p(x)$ is the function of probability density relative to a 
dimensionless variable $x$; $\ln{1 / p(x)}$ is the mean information value; 
$k_B$ is the Boltzmann constant.

Substituting the soliton-like solution (\ref{eq46}) into (\ref{eq49}) we obtain

\begin{equation}
S = -2 k_B \int \limits_0 ^{\infty} \sec {h^2 (\alpha \xi) \ln {\left[ \sec {h^2 (\alpha \xi)} \right] d \left( \frac{x}{u \tau_{\beta}} \right)}} = 4 k_B \int \limits _0 ^{\infty} \frac{\ln \cosh {(\beta y)}}{\cosh ^2{(\beta y)}} dy,
\label{eq50}
\end{equation}

\noindent where $\beta = \alpha (u \tau_{\beta})$ is another (dimensionless) 
scaling factor, $\tau_{\beta}$ is the proper $\beta$-decay time of the active 
component of the nuclear fuel. Let us point out the procedure of making the $x$
argument dimensionless in (\ref{eq50}), which takes into account the fact that 
the neutron flux amplitude is proportional to the phase velocity of the nuclear 
burning wave (see (\ref{eq44})), i.e. $\varphi_m \sim u$. Calculating the 
integral (\ref{eq50}) leads to the following quite simple expression for the 
entropy

\begin{equation}
S = \frac{4 (1 - \ln {2})}{\beta} = 4 ( 1 - \ln {2}) \frac{k_B}{\alpha u \tau_{\beta}} = \frac{4 k_B (1 - \ln {2})}{\sqrt{k_{GB} - k_0}} \frac{2L}{u \tau_{\beta}},
\label{eq51}
\end{equation}

\noindent that in the case of

\begin{equation}
S \sim k_B \frac{2L}{u \tau_{\beta}} = const,
\label{eq52}
\end{equation}

\noindent points to an isentropic transport of the nuclear burning wave.

\begin{figure}
  \begin{center}
    \includegraphics[width=15cm]{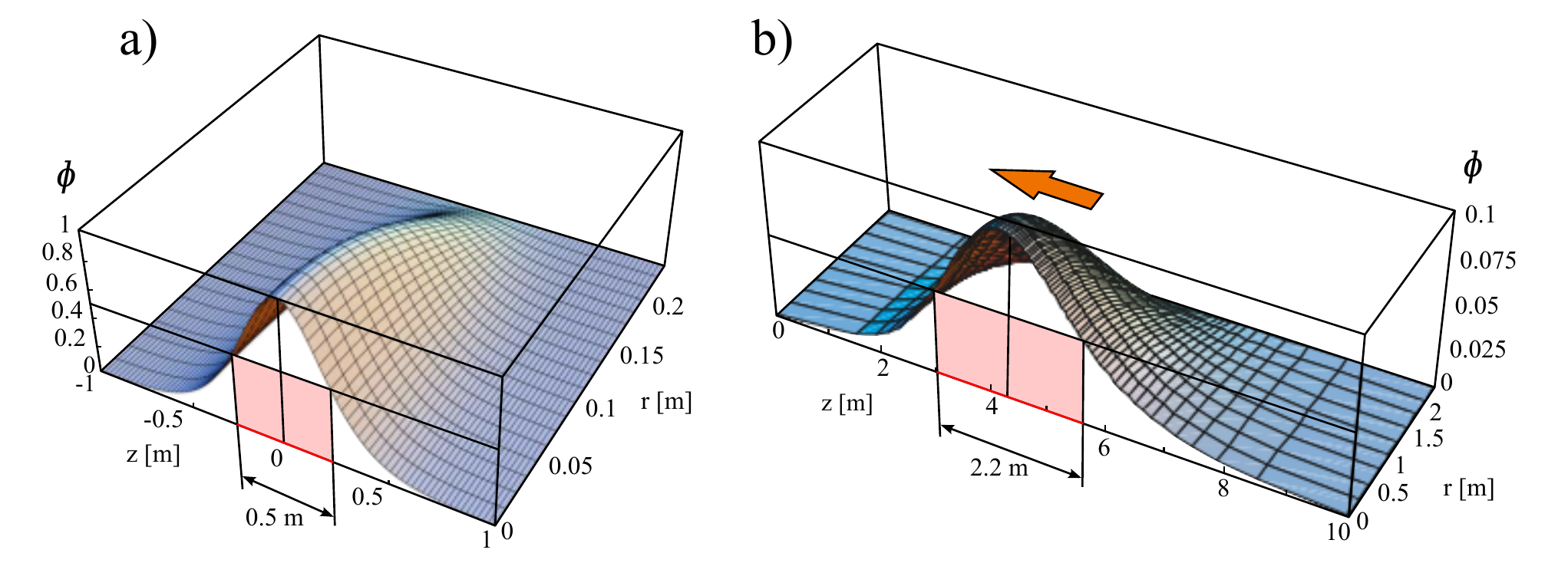}
    \caption{(a) Weakly dispersive wave pattern, obtained by Chen and Maschek \cite{ref12} investigating a 3D-model by van Dam using perturbation method. Example with the following parameters: $L = 0.02~m$, $k_{GB} = 1.04$, $k_0 = 1.02$ and $k_{max} = 1.06$; $\phi_m = 10^{17}~m^{-2}s^{-1}$, $u = 0.244~cm/day$; (b) Strongly dispersive wave pattern, obtained \colorbox{red}{  } Chen et.~al \cite{ref18a} within a 3D-model of traveling wave reactor. Example with the following parameters: $L = 0.017~m$, $k_{GB} = 1.00030$, $k_0 = 0.99955$; $\phi_m = 3 \cdot 10 ^{15} ~m^{-2}s^{-1}$, $u = 0.05~cm/day$.}
    \label{fig3}
  \end{center}
\end{figure}

It is interesting to note here that if the width $\Delta_{1/2} \rightarrow 0$, 
then due to isoentropicity of the process (\ref{eq52}), the form of the soliton 
becomes similar to the so-called Dirac $\delta$-function. Introducing two 
characteristic sizes or two length scales ($l_1 = 1 / \alpha$ and 
$l_2 = \varphi_m$), it is possible to see, that when the first of them is small,
the second one increases and vice versa. It happens because the area $A_{area}$
(\ref{eq42}) under the soliton must remain constant, since 
$A_{area} \propto l_1 l_2$. In this case the soliton entropy tends to zero 
because the entropy is proportional to the ratio of these values 
($S \propto l_1 / l_2 \rightarrow 0 $). These features are the consequence of 
the fact that the scale $l_1$ is a characteristic of a dispersion of the 
process, while another scale $l_2$ is a characteristic of the soliton 
non-linearity. If $l_1 \ll l_2$, then the process is weakly dispersed 
(fig.~\ref{fig3}a). If $l_1 \gg l_2$, then the process is strongly dispersed 
(fig.~\ref{fig3}b). In the latter case the soliton amplitude becomes relatively 
large (a case of $\delta$-function). And finally, if $l_1 = l_2$, then the 
soliton speed $u \propto (l_1^2 l_2^2)^{1/2}$ (see eq.~(\ref{eq45})) is 
proportional to the geometrical mean of the dispersion and non-linearity 
parameters.

On the other hand, it is clear that according to (\ref{eq42}) and (\ref{eq44}), 
the burning wave width $\Delta _{1/2}$ is a parameter that participates in 
formation of the time for neutron fluence accumulation $\tau_{\beta}$ on the 
internal surface of the TWR long fuel rod cladding material

\begin{equation}
\psi_m \sim \frac{\Delta_{1/2}}{u} \varphi_m = \tau_{\varphi} \cdot \varphi_m.
\label{eq53}
\end{equation}

\begin{figure}
  \begin{center}
    \includegraphics[width=10cm]{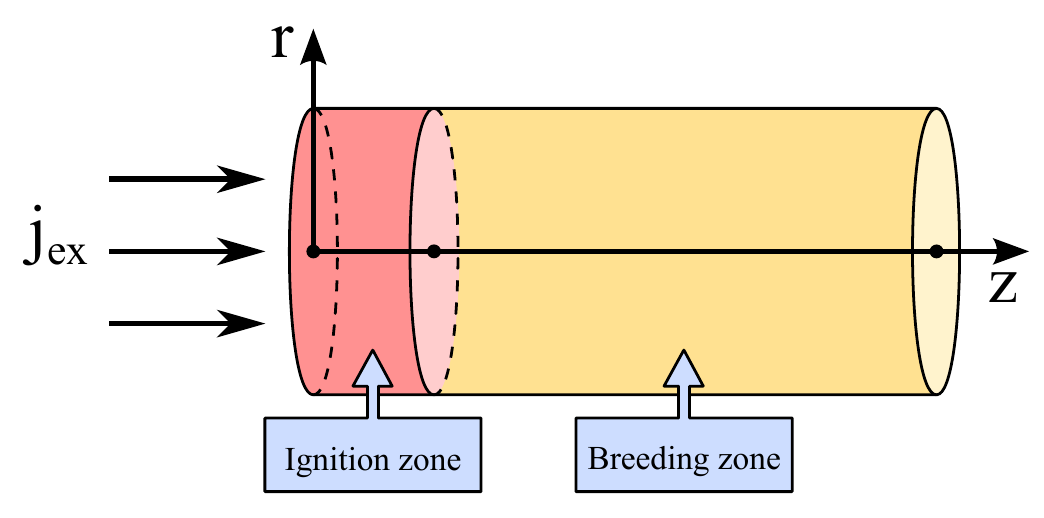}
    \caption{Schematic sketch of the two-zone cylindrical TWR.}
    \label{fig4}
  \end{center}
\end{figure}

And finally, one more important conclusion. From the analysis of (\ref{eq42}) 
and (\ref{eq43}) it is clear that the initial parameter $k_0$ for burning zone 
(fig.~\ref{fig4}) is predefined solely by the nuclear burning wave burn-up 
conditions, i.e. by the parameters of an external neutron source and burn-up 
area composition (fig.~\ref{fig4}). In other words, it means that by tuning the 
corresponding burn-up conditions for a given nuclear fuel composition, we can 
set the certain value of the nuclear burning front width. Moreover, by selecting
the corresponding equilibrium $\tilde{n}_{fis}$ and critical $n_{crit}$ 
concentrations of the active nuclear fuel component, we can define the required 
value of the nuclear burning wave speed $u$. Hence an obvious way for us to 
control the corresponding neutron fluence $\tau_{\varphi}$ accumulation time in 
the cladding material of the TWR fuel rod.

Hence we can make an important conclusion that the realization of the TWR with
inherent safety requires the knowledge about the physics of nuclear burning wave
burn-up and the interrelation between the speed of nuclear burning wave and the 
fuel composition. As is shown in \cite{ref29}, the properties of the fuel are 
completely defined by  the equilibrium $\tilde{n}_{fis}$ and critical $n_{crit}$
(see (\ref{eq3a})) concentrations of the active nuclear fuel component. We 
examine this in more detail below.

\section{Control parameter and condition of existence of stationary wave of nuclear burning.}
\label{sec3}

The above stated rises a natural question: "What does the nuclear burning wave 
speed in uranium-plutonium (\ref{eq1a}) and thorium-uranium (\ref{eq2a}) cycles 
mainly depend on?" The answer is rather simple and obvious. The nuclear burning 
wave speed in both cycles (far away from the burn-up source) is completely 
characterized by its equilibrium $\tilde{n}_{fis}$ and critical $n_{crit}$ 
concentrations of the active fuel component.

\begin{figure}
  \begin{center}
    \includegraphics[width=10cm]{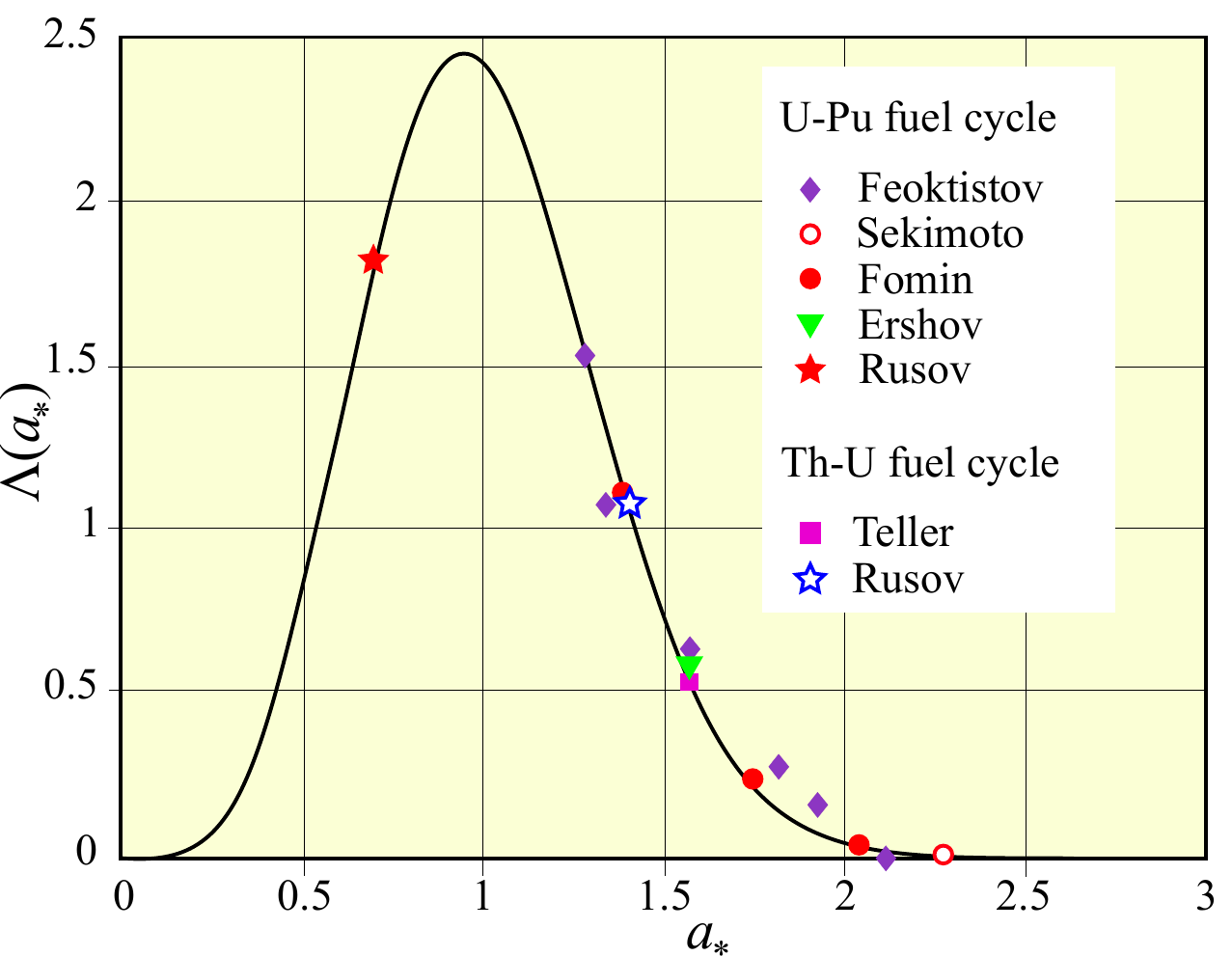}
    \caption{The theoretical (solid line) and calculated (points) dependence of 
             $\Lambda (a_{*}) = u \tau_{\beta} / 2L$ on the parameter $a$ 
	     \cite{ref29}.}
    \label{fig5}
  \end{center}
\end{figure}

First of all, this is determined by a significant fact that the equilibrium 
$\tilde{n}_{fis}$ and critical $n_{crit}$ concentrations of the active fuel 
component completely identify the neutron-multiplying properties of the fuel 
environment. They are the conjugate pair of the integral parameters, which due 
to their physical content, fully and adequately characterize all the physics of 
the nuclear transformations predefined by the initial fuel composition. This is 
also easy to see from a simple analysis of the kinetics equations solutions for 
the neutrons and nuclei, used in different models \cite{ref5,ref6,ref7,ref8,
ref9,ref12,ref14,ref16,ref18,ref18a,ref19,ref20,ref21,ref22,ref23,ref26,ref27,
ref28,ref29}. It mean that regardless of the nuclear cycle type and initial 
fuel composition, the nuclear burning wave speed is defined by the equilibrium 
$\tilde{n}_{fis}$ and critical $n_{crit}$ concentrations of the active fuel 
component through the so-called para-parameter $a$ (see (\ref{eq3a})). 
Consequently, as the numerical simulation results show, 
(fig.~\ref{fig5}~\cite{ref29}), it follows the Wigner statistics.

This is corroborated by the following fact. The papers \cite{ref20,ref21} study 
a boundary-value problem for the stationarity of the nuclear burning wave, 
formulated within the diffusion equation for the neutron fluence and the 
kinetics equations for the nuclear density of the mother and daughter nuclides:

\begin{equation}
\frac{\partial \psi}{\partial t} = \nu \cdot D \cdot \Delta \psi + G \left( \vec{N}, \psi \right),
\label{eq54}
\end{equation}

\begin{equation}
\frac{\partial \vec{N}}{\partial t} = \widehat{\sigma} \cdot \vec{N} \cdot \frac{\partial \psi}{\partial t} - \hat{\lambda} \vec{N},
\label{eq55}
\end{equation}

\noindent where

\begin{equation}
G\left( \vec{N}, \psi \right) = \int \limits_0^{\psi_{end}} g \left( \vec{N} \right) d \psi
\label{eq56}
\end{equation}

\noindent is a function\footnote{The expression (\ref{eq55}) should be 
understood as an integral along the system path in configuration space of the 
variables ($N$,$\psi$) for the given spatial point.} of neutron fluence 
generation, $g(\vec{N})$ is the neutron generation function, which is a linear 
function of the nuclei concentration $N_i$, $\vec{N}$ is the column of the 
mother and daughter nuclides $N_i$, $\psi$ is the neutron fluence, $\psi_{end}$ 
is some final neutron fluence, $\nu$ is the average neutron speed, $D$ is the 
diffusion coefficient, $\hat{\sigma}$ is the matrix of the microscopic neutron 
absorption and capture cross-sections, and $\hat{\lambda}$ is the matrix of the 
radioactive decay constants for the $\beta$-active nuclei.

It is shown \cite{ref20}, that this boundary-value problem for the equations 
(\ref{eq54})-(\ref{eq55}) is a spectral non-linear differential problem. A 
dimensionless speed $W$ of the nuclear burning wave was chosen as a spectral 
(free) parameter. Such representation of the problem makes it possible to 
investigate the final fluence behavior caused by the change of the absorbent 
properties, whose concentration, according to \cite{ref20,ref21,ref26,ref27}, 
controls (in a zero approximation of perturbation theory) the value of the 
dimensionless wave speed:

\begin{equation}
W = \frac{u \tau_2}{L} = \frac{1}{b} \left( p_0 - p \right), ~~ at ~ W \ll 1,
\label{eq57}
\end{equation}

\noindent where $u$ is the nuclear burning wave speed, $\tau_2$ is the internal 
time scale, equal to the characteristic time of the intermediate nuclide 
$\beta$-decay ($\tau_2 = 3.47$ days for $^{239}Np$ in U-Pu cycle and 
$\tau_2 = 36.6$ days for $^{233}Pa$ in Th-U cycle), $L$ is the neutron diffusion
length, $1/b$ is the linear dependency slope (\ref{eq57}), $p$ is the 
dimensionless effective concentration of the absorbent, $p_0$ is the upper limit
of the absorbent concentration the nuclear wave can exist for.

Here we obtain an important result showing that the final fluence and the 
absorbent concentration in the limiting case $W \rightarrow 0$ (zero order of 
perturbation theory) are the solutions of the two equilibrium conditions for a 
stationary nuclear burning wave:

\begin{equation}
\int \limits _0 ^{\psi_{end}} g d\psi = 0,
\label{eq58}
\end{equation}

\begin{equation}
M \left( \psi_{end} \right) = \int \limits_0 ^{\psi_{end}} \left( \psi_{end} - \psi \right) g d\psi = 0.
\label{eq59}
\end{equation}

For the sake of the more clear understanding of the physical sense of these 
conditions, the authors of \cite{ref20} suggest a simple, yet elegant and deep 
analogy. The obtained conditions exactly coincide by form with the conditions 
for a lever subject to a distributed perpendicular force $g(\psi)$ applied on a 
segment $0 \leqslant \psi \leqslant \psi_{end}$ along the lever. In other words,
according to this analogy, the conditions (\ref{eq58}) and (\ref{eq59}) 
represent the conditions of the zero total force and total momentum 
respectively. Therefore, if the first condition is an integral condition of the 
neutrons generation and absorption equality, then by analogy the second 
condition may be called the condition of the neutrons generation and absorption 
"momenta" equality.

If the expression for $g \left( \vec{N} \right)$ is presented in the form of a 
sum of the fuel $g^F$ and absorbent $g^A = p$ contributions in the neutron 
generation:

\begin{equation}
g = g^F - g^A,
\label{eq60}
\end{equation}

\noindent then according to the conditions (\ref{eq58}) and (\ref{eq59}), it is 
easy to obtain a modified condition (\ref{eq59}) for the mean "momenta" of the 
neutrons generation $\left \langle M^F \right \rangle$ and absorption 
$\left \langle M^A \right \rangle$ in the form~\cite{ref26}

\begin{equation}
\left \langle M^F \left( \psi_{end} \right) \right \rangle = \frac{\int \limits_0^{\psi_{end}} g^F \psi d \psi}{\int \limits _0 ^{\psi_{end}} g^F d\psi} = \frac{\int \limits_0^{\psi_{end}} g^A \psi d \psi}{\int \limits _0 ^{\psi_{end}} g^A d\psi} = \left \langle M^A \left( \psi_{end} \right) \right \rangle,
\label{eq61}
\end{equation}

\noindent which by definition (\ref{eq58}) has a trivial root 
$\psi = \psi_{end}$.

\begin{figure}
  \begin{center}
    \includegraphics[width=15cm]{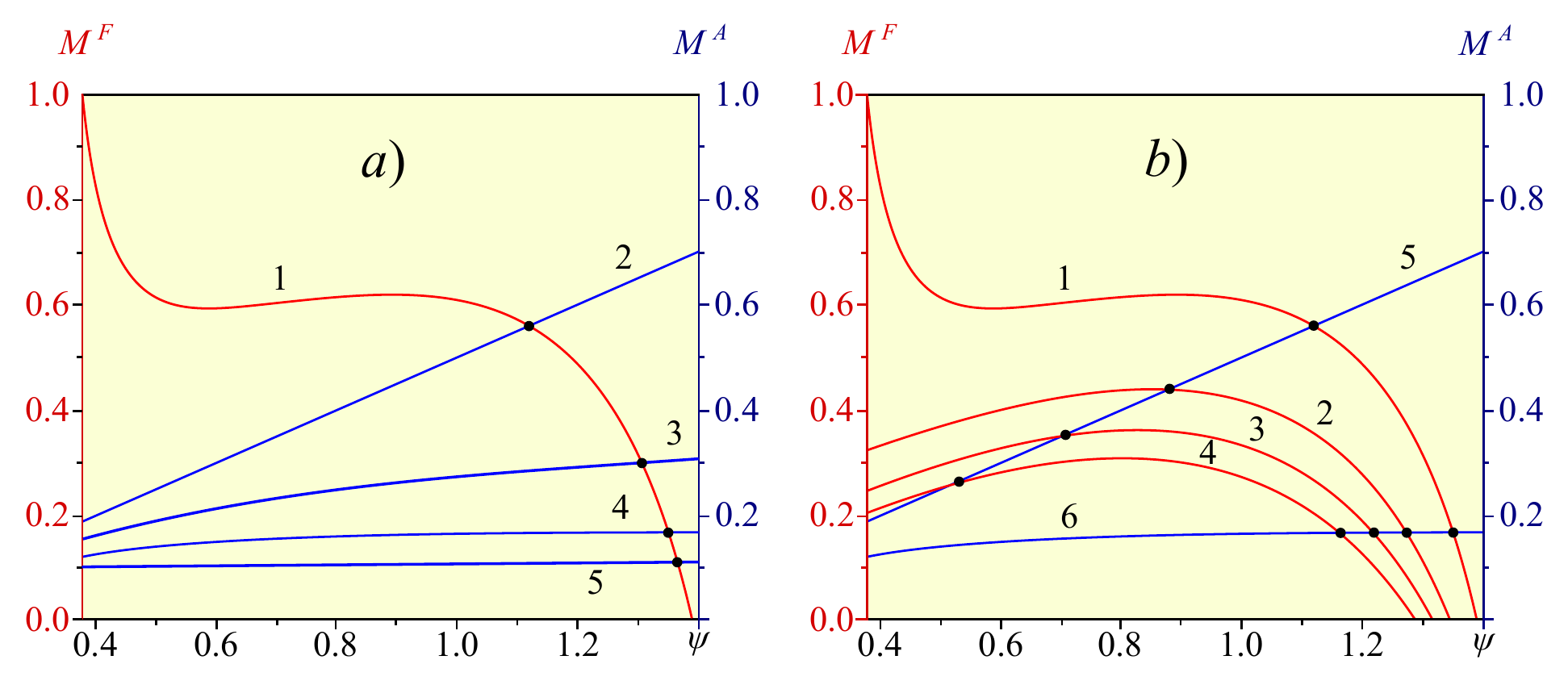}
    \caption{Graphical solution of the equation (\ref{eq61}) for the final 
             fluence $\psi_{end}$ with (a) different speeds of the absorbent 
	     burn-out and (b) different enrichment of the $^{239}Pu$ for the 
	     non-burnable and burnable adsorbents. At fig.\ref{fig6}a the curve 
	     1 is a graph of the right part of the equation, and the curves 2-5 
	     represent the left part for the absorbent cross-sections of 0, 1, 2
	     and 3 barn respectively. At fig.\ref{fig6}b the curves 1-4 are 
	     the plots of the right part for the initial $^{239}Pu$ 
	     concentrations of 0, 3, 5, 7\% respectively, and the curves 5 and 6
	     represent the left part of the equation for the absorbent 
	     cross-sections of 0 and 2 barn respectively. Adopted from 
	     \cite{ref26}.}
    \label{fig6}
  \end{center}
\end{figure}

The main advantage of such presentation of the equation (\ref{eq61}) is that 
its left part, i.e. $\left \langle M^F \right \rangle$, is defined solely by 
the properties of the fuel, while its right part, i.e. 
$\left \langle M^A \right \rangle$, is defined solely by the absorbent. 
Moreover, according to \cite{ref26}, the initial absorbent concentration is 
absent in the equation (\ref{eq61}). The plots for left and right parts of the 
equation (\ref{eq61}) are presented at fig.~\ref{fig6}~\cite{ref26} for 
different cases.

\begin{figure}
  \begin{center}
    \includegraphics[width=15cm]{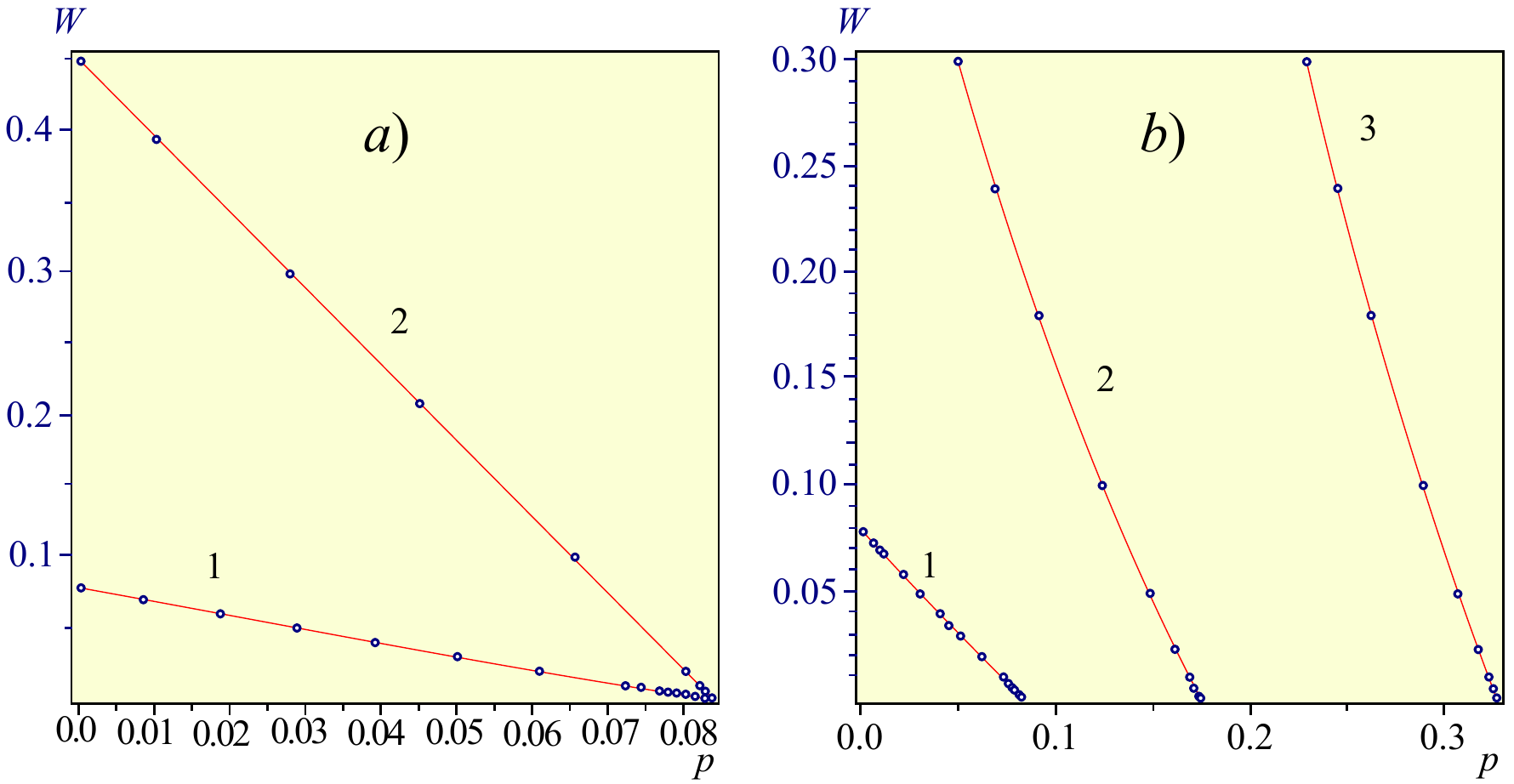}
    \caption{Dependence of the nuclear burning wave speed on the non-burnable 
             absorbent concentration $p$, (a) taking into account the burn-out 
	     of the $^{239}Np$ (curve 1) and not taking into account the 
	     burn-out of the $^{239}Np$ (curve 2), (b) for the different initial
	     fuel compositions: 1 - 100\% $^{238}U$; 2 - 93\% $^{238}U$ + 7\% 
	     $^{235}U$; 3 - 93\% $^{238}U$ + 7\% $^{239}Pu$. Adopted from 
	     \cite{ref26}.}
    \label{fig7}
  \end{center}
\end{figure}

A simple analysis of the fig.~\ref{fig6} shows that the increase of the 
non-burnable absorbent concentration (curves 2-5 at fig.~\ref{fig6}a) leads to
the final $\psi_{end}$ fluence increase, but according to (\ref{eq57}) -- to the
nuclear burning wave speed decrease at the same time. And vice versa, with the
non-burnable absorbent concentration increase (curves 1-4 at fig.~\ref{fig6}b) 
the value of the final fluence $\psi_{end}$ decreases, while the velocity of the
nuclear burning wave at $p=0$ also decreases in a complex manner (see 
fig.~\ref{fig7}), depending on the slope behavior in (\ref{eq57}). The increase 
in the absorbent concentration (curves 5-6 at fig.~\ref{fig6}b) does not change 
the effect qualitatively, but leads to a strong quantitative change. The final 
fluence turns out to be very sensible to the burnable absorbent concentration: 
curve~6 at fig.~\ref{fig6}b for the burnable absorbent crosses the curves 1-4 at
much greater values of final fluence.

Thus, basing on the numerical solution of the boundary-value problem for the 
stationary nuclear burning wave and the developed earlier analytical theory of 
the nuclear burning wave~\cite{ref20,ref21}, it was shown in the 
papers~\cite{ref26,ref27} (in a zero order of perturbation theory for $W$) that 
the control parameter, which makes it possible to change the speed of nuclear 
burning and hence the reactor power, is the effective concentration of the 
absorbent. Moreover, this control parameter allows both to increase and 
decrease the mentioned parameters. According to \cite{ref20,ref21,ref26,ref27}, 
"\dots this is achieved by a purposeful change of the initial reactor 
composition".

The conclusion made by the authors of \cite{ref20,ref21,ref26,ref27} requires 
the following comment. According to our earnest conviction, the results of these
works are indeed very interesting and informative, but the conclusion are not
satisfactory however strange it may seem. And here is why.

As a matter of fact, as it was shown in \cite{ref29}, the control parameter is 
not the effective absorbent concentration ($p$) (or the nuclear burning wave 
speed ($W$), or the maximal neutron fluence ($\psi_{end}$)), but the so-called 
\textbf{para-parameter} of the nuclear TWR burning. As it is shown above (see 
(\ref{eq3a})), this para-parameter is formed by a conjugated pair of the 
integral parameters, i.e. by the equilibrium and critical concentrations of the 
active nuclear fuel component. It is important to note that each of this 
concentrations varies during the nuclear burning, but their ratio

\begin{equation}
a = \frac{\pi}{2} \sqrt{\frac{n_{crit}}{\tilde{n}_{fis} - n_{crit}}}
\label{eq62}
\end{equation}

\noindent is a characteristic constant value for the given nuclear burning 
process \cite{ref29}. In addition to that this para-parameter also determines 
(and it is extremely important!) the conditions for the nuclear burning wave 
existence (\ref{eq3a}), the neutron nuclear burning wave speed (see 
(\ref{eq3a})) and the dimensionless width (\ref{eq62}) of the super-critical 
area in the burning wave of the active nuclear fuel component.

Therefore, when the authors \cite{ref20,ref21,ref26,ref27} state that they 
control the values of the parameters with purposeful variation of the initial 
reactor composition, it actually means that changing the effective concentration
of the absorbent, they purposefully and definitely change (by definition (see 
\cite{ref3,ref29})) the equilibrium and critical concentrations of the active 
component, i.e. the para-parameter (\ref{eq62}) of the nuclear TWR-burning 
process.

It is necessary to note that the adequate understanding of the control parameter
determination problem is not a simple or even a scholastic task, but is 
extremely important for the effective solution of another problem (the major 
one, in fact!), related to investigation of the nuclear burning wave stability 
conditions. In our opinion, the specified conditions for the stationary nuclear 
burning wave existence (\ref{eq3a}) and (\ref{eq58})-(\ref{eq59}) obtained 
in the papers \cite{ref29} and \cite{ref20,ref21,ref26} respectively, reveal the
path to a sensible application of the so-called direct Lyapunov method 
\cite{ref39} (the base theory for the movement stability), and thus the path to 
a reliable justification of the Lyapunov functional minimum existence (if it 
does exist) \cite{ref39,ref40,ref41,ref42}. Some variants of a possible solution
stability loss due to anomalous evolution of the nuclear fuel temperature are 
considered in Chapter~\ref{sec5}.

At the same time one may conclude that the "differential"~\cite{ref29} and 
"integral"~\cite{ref20,ref21,ref26} conditions for the stationary nuclear 
burning wave existence provide a complete description of the wave reactor 
physics and can become a basis for the future engineering calculations of a 
contemporary TWR project with the optimal or preset wave properties.

\section{On the dependence of the damaging dose on neutron fluence, phase 
         velocity and dispersion of the solitary burn-up waves}
\label{sec4}

As follows from the expression for the soliton-like solution (\ref{eq46a}), it 
is defined by three parameters -- the maximal neutron flux $\psi_m$, the phase 
$\alpha$ and the speed $u$ of nuclear burning wave. And even if we can control 
them, it is still unclear, which condition determines the optimal values of 
these parameters. Let us try to answer this question shortly.

\begin{figure}
  \begin{center}
    \includegraphics[width=10cm]{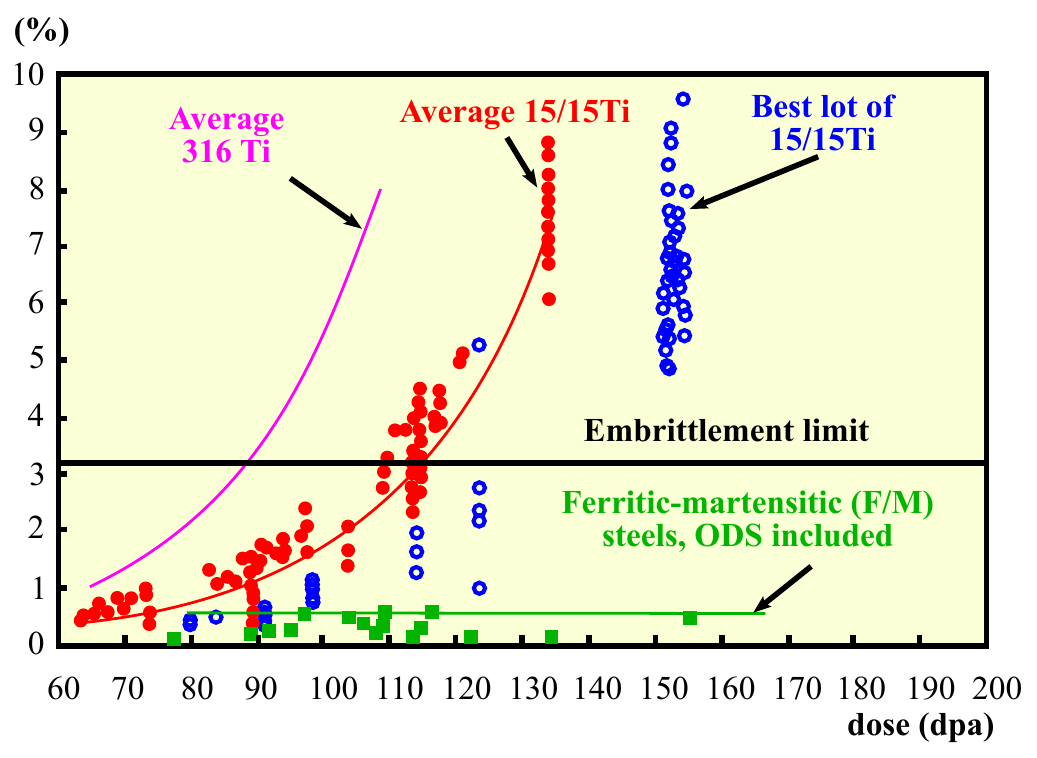}
    \caption{Swelling of austenitic Phenix cladding compare to ferritic-martensitic materials, ODS included. Adopted from \cite{ref48}.}
    \label{fig8}
  \end{center}
\end{figure}

It is known that a high cost effectiveness and competitiveness of the fast 
reactors, including TWR, may be achieved only in case of a high nuclear fuel 
burn-up\footnote{Since the maximum burn-up of the FN-600 reactor is currently 
$\sim 10\%$~\cite{ref43}, the burn-up degree of $\sim 20\%$ for the TWR may be 
considered more than acceptable.}. As the experience of the fast reactors 
operation shows, the main hindrance in achieving the high nuclear fuel burn-up 
is the insufficient radiation resistance of the fuel rod shells. Therefore the 
main task of the radiation material science (along with the study of the physics
behind the process) is to create a material (or select among the existing 
materials), which would keep the required level of performance characteristics 
being exposed to the neutron irradiation. One of the most significant phenomena 
leading to a premature fuel rods destruction is the vacansion swelling of the 
shell material~\cite{ref44,ref45,ref46,ref47}. Moreover, the absence of the 
swelling saturation at an acceptable level and its acceleration with the 
damaging dose increase leads to a significant swelling (volume change up to 30\%
and more) and subsequently to a significant increase of the active zone elements
size. The consequences of such effect are amplified by the fact that the high 
sensitivity of the swelling to temperature and irradiation damaging dose leads 
to distortions of the active zone components form because of the temperature and
irradiation gradients. And finally, one more aggravating consequence of the high
swelling is almost complete embrittlement of the construction materials at 
certain level of swelling\footnote{It is known that the fuel rod shell diameter 
increase due to swelling is accompanied by an anomalously high corrosion damage 
of the shell by the fuel \cite{ref43}}. Consequently, in order to estimate the 
possible amount of swelling, a damaging dose (measured in dpa) initiated by the 
fast neutrons, e.g. in the fuel rod shells, must be calculated 
(fig.~\ref{fig8}).

Usually in order to evaluate the displacements per atom (DPA) created by the 
spallation residues, the so-called modified NRT method is applied 
\cite{ref49,ref50}, which takes into account the known Lindhard correction 
\cite{ref51}. Within such modified NRT method, the total number of 
displacements produced by the residues created by spallation reactions in the 
energetic window can be calculated as the addition of the displacements 
produced by each of these residues ($Z$,$A$), that in its turn leads to the 
following expression for the displacements per atom:

\begin{equation}
n_{dpa} = t \cdot \sum \limits _i ^N \left \langle \sigma_{dpa}^i \right \rangle \int \limits_{E_i - 1}^{E_i} \Phi \left( E_i \right) d E_i = \sum \limits _{i=1} ^N \left \langle \sigma_{dpa}^i \right \rangle \varphi_i t,
\label{eq63}
\end{equation}

\noindent where

\begin{equation}
\left \langle \sigma_{dpa}^i \right \rangle \left( E_i \right) = \left \langle \sigma_d \left( E_i, Z, A \right) \right \rangle \cdot d \left( E_d \left( Z,A \right) \right),
\label{eq64}
\end{equation}

\noindent while $\sigma_d \left( E_i, Z, A \right)$ is the displacement 
cross-section of recoil atom ($Z$,$A$), produced at incident particle energy 
$E_i$, $\Phi (E)$ is the energy-dependent flux of incident particles during 
time $t$ and $d \left( E_d (Z,A) \right)$ is the number of displacements created
at threshold displacement energy $E_d$ of recoil atom ($Z$,$A$) or its so-called
damage function.

Actually, all the recoil energy of the residue $E_r$ is not going to be useful 
to produce displacements because a part of it is lost inelastic scattering with 
electrons in the medium. An estimation of the damage energy of the residue can 
be calculated using the Lindhard factor $\xi$~\cite{ref51}

\begin{equation}
E_{dam} = E_r \xi.
\label{eq65}
\end{equation}

The number of displacements created by a residue ($Z$,$A$) are calculated using 
this damage energy (\ref{eq65}) and the NRT formula:

\begin{equation}
d \left( E_d (Z,A) \right) = \eta \frac{E_{dam}}{2 \left \langle E_d \right \rangle},
\label{eq66}
\end{equation}

\noindent where $\left \langle E_d \right \rangle$ is the average threshold 
displacement energy of an atom to its lattice site, $\eta = 0.8$~
\cite{ref49,ref50}.

\begin{figure}
  \begin{center}
    \includegraphics[width=10cm]{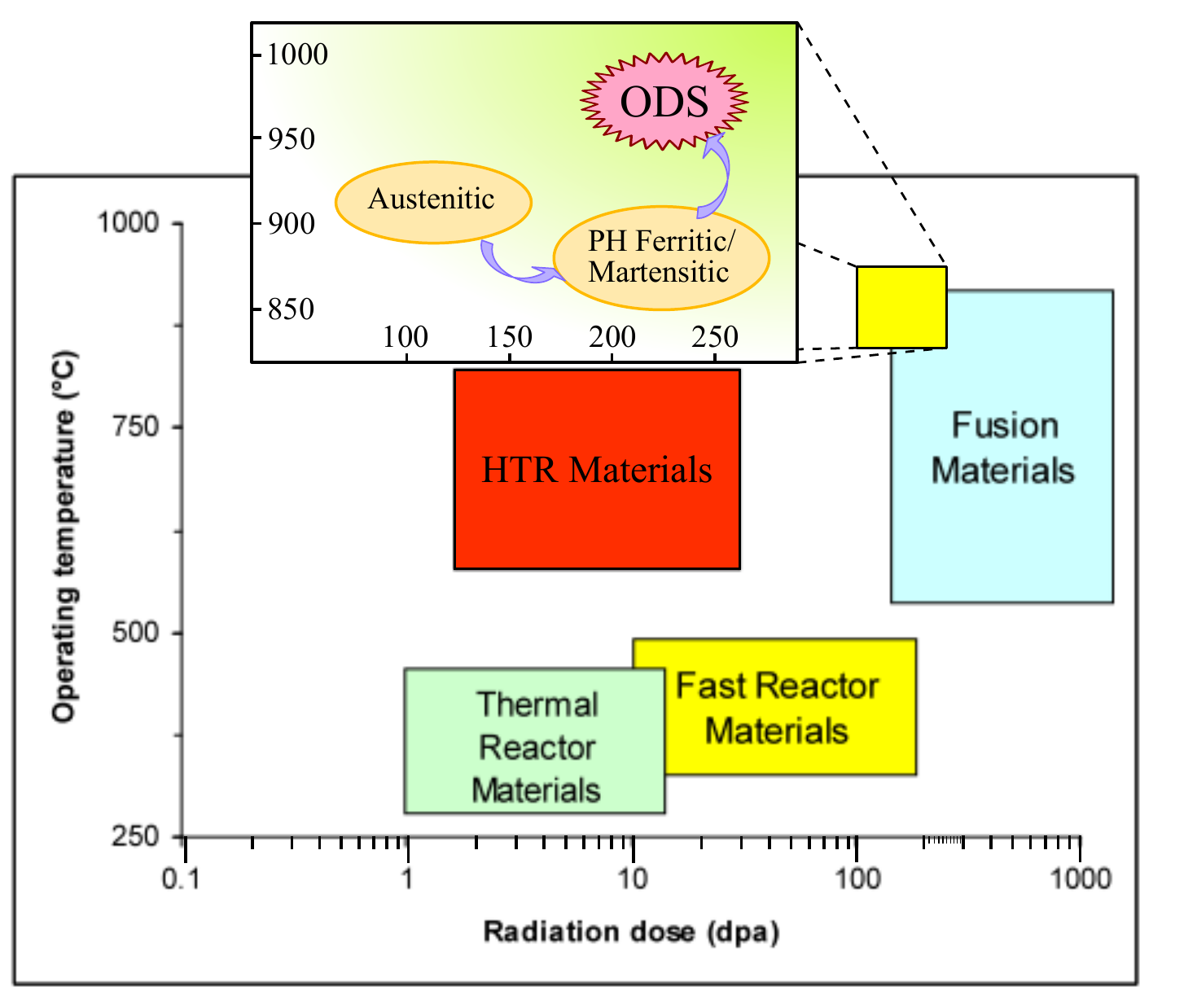}
    \caption{Operating condition for core structural materials in different 
             power reactors \cite{ref52}. The upper yellow inset represents the 
	     data of Pukari~M. \& Wallenius \cite{ref48}.}
    \label{fig9}
  \end{center}
\end{figure}

Consequently the condition of the maximal damaging dose for the cladding 
materials of the fast neutron reactors, taking into account the  metrological 
data of IAEA~\cite{ref52} (see fig.~\ref{fig9}) and contemporary estimates by 
Pukari~M. \& Wallenius (see yellow inset at fig.~\ref{fig8}) takes the form:

\begin{equation}
n_{dpa} \simeq \left \langle \sigma_{dpa} \right \rangle \cdot \varphi \cdot \frac{2 \Delta_{1/2}}{u} \leqslant 200 ~~[dpa].
\label{eq67}
\end{equation}

In this case the selection strategy for the required wave parameters and allowed
values of the neutron fluence for the future TWR project must take into account 
the condition for the maximal damaging dose (\ref{eq66}) for cladding materials 
in fast neutron reactors, and therefore, must comply with the following 
dpa-relation:

\begin{equation}
\left \langle \sigma_{dpa} \right \rangle \cdot \varphi \cdot \frac{\Delta_{1/2}}{u} \leqslant 100 ~~[dpa].
\label{eq68}
\end{equation}

\begin{figure}
  \begin{center}
    \includegraphics[width=10cm]{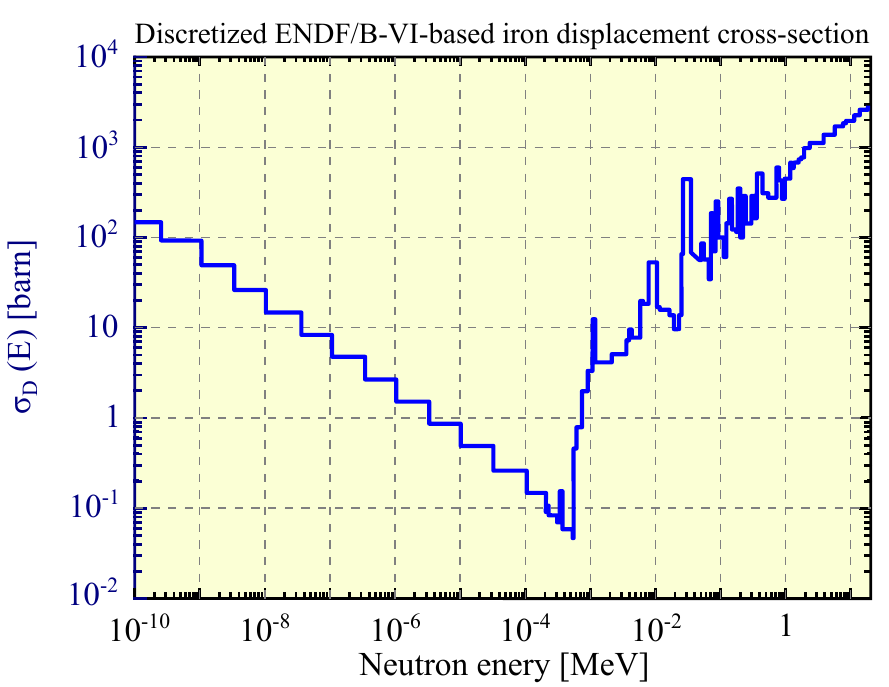}
    \caption{Discretized displacement cross-section for stainless steel based on
             the Lindhard model and ENDF/B scattering cross-section. Adopted 
	     from \cite{ref53}.}
    \label{fig10}
  \end{center}
\end{figure}

The question here is whether or not the parameters of the wave and neutron 
fluence which can provide the burn-up level of the active nuclear fuel component
in TWR-type fast reactor of at least 20\% are possible. Since we are interested 
in the cladding materials resistible to the fast neutron damaging dose, we shall
assume that the displacement cross-section for the stainless steel, according to
Mascitti~et~al.~\cite{ref53} for neutrons with average energy 2~MeV equals 
$\left \langle \sigma_{dpa} \right \rangle \approx 1000~dpa$ (fig.~\ref{fig10}).

The analysis of the nuclear burning wave parameters in some authors' models of 
TWR \cite{ref4,ref5,ref13,ref14,ref15,ref18a,ref25,ref26,ref28,ref29} presented 
in Table~\ref{tab1}, shows that in the case of U-Pu cycle none of the considered
models satisfy the dpa-parameter, while two Th-U cycle models by Teller~
\cite{ref4} and Seifritz~\cite{ref5} groups correspond well to the major 
requirements to wave reactors.

\begin{table}
\rowcolors{3}{lightgray}{white}
\begin{tabular}{|l|c|c|c|c|c|c|c|c|}
  \hline
  & $\Delta_{1/2}$ & $u$         & $\varphi$ & $\psi$ & $\left \langle \sigma_{dpa} \right \rangle$ & \multirow{2}{*}{$\dfrac{n_{dpa}}{200}$} & Fuel & \multirow{2}{*}{Solution} \\
  & [cm]                  & [cm/day] & [cm$^{-2}$s$^{-1}$] &  [cm$^{-2}$] & [barn] & &burn-up & \\
  \hline
  \multicolumn{9}{|c|}{\textbf{U-Pu cycle}} \\
  \hline
  Sekimoto \cite{ref14}  & 90 & 0.008 & 3.25 $\cdot$ 10$^{15}$ & 3.2 $\cdot$ 10 $^{23}$ & 1000 & 3.2 & $\sim$43\% & No \\
  \hline
  Rusov \cite{ref29} & 200 & 2.77 & 10$^{18}$ & 6.2$\cdot$10$^{24}$ & 1000 & 62 & $\sim$60\% & No\\
  \hline
  Pavlovich \cite{ref26} & -- & 0.003 & -- & 1.7$\cdot$10$^{24}$ & 1000 & 17 & $\sim$30\% & No\\
  \hline
  Fomin \cite{ref15} & 100 & 0.07 & 2 $\cdot$ 10$^{16}$ & 2.5$\cdot$10$^{24}$ & 1000 & 25 & $\sim$30\% & No\\
  \hline
  Fomin \cite{ref13} & 125 & 1.7 & 5 $\cdot$ 10$^{17}$ & 3.2$\cdot$10$^{24}$ & 1000 & 32 & $\sim$40\% & No\\
  \hline
  Chen \cite{ref18} & 216 & 0.046 & 3 $\cdot$ 10$^{15}$ & 1.2$\cdot$10$^{24}$ & 1000 & 12 & $\sim$30\% & No\\
  \hline
  T.~Power \cite{ref28} & -- & -- & -- & -- & -- & 1.75 & $\sim$20\% & No\\
  \hline
  \multicolumn{9}{|c|}{\textbf{Th-U cycle}} \\
  \hline
  Teller \cite{ref4} & 70 & 0.14 & $\sim$2 $\cdot$ 10$^{15}$ & 8.6$\cdot$10$^{22}$ & 1000 & 0.96 & $\sim$50\% &  {\color{red} Yes}\\
  \hline
  Seifritz \cite{ref5} & 100 & 0.096 & 10$^{15}$ & 9.0$\cdot$10$^{22}$ & 1000 & 0.90 & $\sim$30\% &  {\color{red} Yes}\\
  \hline
  Melnik \cite{ref25} & 100 & 0.0055 & 0.5$\cdot$10$^{16}$ & 7.9$\cdot$10$^{24}$ & 1000 & $\sim$80 & $\sim$50\% & No \\
  \hline
  \multicolumn{9}{|c|}{\textbf{U-Pu (+ moderator)}} \\
  \hline
  Example & 100 & 0.234 & 2.5 $\cdot$ 10$^{15}$ & 9.2$\cdot$10$^{23}$ & 100 & 0.92 & $\sim$20\% &  {\color{red} Yes}\\
  \hline
  Ideal TWR & -- & -- & -- & 10$^{24}$ & 100 & 1.0 & $\geqslant$20\% &  {\color{red} Yes}\\
  \hline
\end{tabular}
\caption{Results of the numerical experiments of the wave mode parameters based 
         on U-Pu and Th-U cycles}
\label{tab1}
\end{table}

On the other hand, the authors of \cite{ref13,ref14,ref15,ref18a,ref25,ref26,
ref28}, obviously did not take the problem of dpa-parameter in cladding 
materials into account, since they were mainly interested in the fact of the 
wave mode of nuclear burning existence in U-Pu and Th-U cycles at the time.

However, as the analysis of Table~\ref{tab1} shows, the procedure of account for
the dpa-parameter is not problematic, but it leads to unsatisfactory results 
relative to the burn-out of the main fissle material.

In other words, from the analysis of Table~\ref{tab1} it follows that when 
$\left \langle \sigma_{dpa} \right \rangle \approx 1000~dpa$, the considered 
above dpa-condition for the maximum possible damaging dose for cladding 
materials of the fast neutron reactors

\begin{equation}
\psi_{1000} = \dfrac{\Delta_{1/2}}{u} \varphi \simeq 10^{23} ~~[cm^{-2}],
\label{eq68a}
\end{equation}

\noindent is not met by any example in Table~\ref{tab1}. Here $\psi_{1000}$ is 
the neutron fluence in case $\left \langle \sigma_{dpa} \right \rangle \approx 
1000~dpa$, $\varphi$ is the neutron flux, $\Delta_{1/2}$ and $u$ are the width 
and speed of the soliton-like nuclear burning wave.

So on the one hand, the neturon fluence must be increased by an order of 
magnitude to increase the burn-up level significantly, and on the other hand, 
the maximum damaging dose for the cladding materials must also  be reduced by an
order of magnitude. Such a controversial condition may be fulfilled considered 
that the reduction of the fuel rod shell radioactive damage for a given amount 
may be achieved by reducing the neutron flux density and energy (see fig.~
\ref{fig10}). The latter is achieved by placing a specially selected substance 
between fissile medium and fuel rod shell, which has the suitable 
characteristics of neutron moderator and absorbent.

At the same time it is known from the reactor neutron physics~\cite{ref35,
ref64}, that the moderator layer width estimate $R_{mod}$ is:

\begin{equation}
R_{mod} \simeq \dfrac{1}{\Sigma_S + \Sigma_a} \cdot \dfrac{1}{\xi} \ln {\dfrac{E_{fuel}}{E_{mod}}},
\end{equation}

\noindent where $\Sigma_S \approx \left \langle \sigma_{S} \right \rangle 
N_{mod}$ and $\Sigma_a \approx \left \langle \sigma_{a} \right \rangle N_{mod}$ 
are the macroscopic neturon scattering and absorption cross-sections 
respectively, $\left \langle \sigma_{S} \right \rangle$ and $\left \langle 
\sigma_{a} \right \rangle$ are the microscopic neutron scattering and absorption
cross-sections respectively averaged by energy interval of the moderating 
neutrons from $E_{fuel} = 2~MeV$ to $E_{mod} = 0.1~MeV$, $N_{mod}$ is the 
moderator nuclei density, $\xi = 1 + (A + 1)^2 \ln {\left[ (A - 1) / A + 1 
\right] / 2A}$ is the neutron energy decrement of its moderation in the 
moderator-absorbent medium with atomic number $A$.

It is clearly seen that the process of neutron moderation from 2.0~MeV to 
0.1~MeV energy in moderator-absorbent of a given width (see Table~\ref{tab2}) 
creates a new, but satisfactory level of maximum possible damaging dose for the 
cladding materials, corresponding to $\left \langle \sigma_{dpa} \right \rangle 
\approx 100~dpa$ (fig.~\ref{fig10}). Therefore if we are satisfied with the main
fissile material burn-out level around $\sim$20\%, then analyzing 
Table~\ref{tab1} and Table~\ref{tab2}, the conditions accounting for the 
dpa-parameter problem and contemporary level of the radioactive material science
will have the following form:

\begin{equation}
\psi_{100} = \dfrac{\Delta_{1/2}}{u} \varphi \simeq 10^{24} ~~[cm^{-2}],
\label{eq68b}
\end{equation}

\noindent where $\psi_{100}$ is the neutron fluence (with 0.1~MeV energy) on 
cladding materials surface in case $\left \langle \sigma_{dpa} \right \rangle 
\approx 100~dpa$ (see fig.\ref{fig10} and "ideal" case in Table~\ref{tab1}).

\begin{table}
  \begin{tabular}{|c|c|c|c|c|c|c|}
    \hline
     \multirow{4}{2cm}{Moderator} &  \multirow{4}{1.5cm}{Mass number, $A$} & \multirow{4}{2cm}{Mean logarithmic energy, $\xi$} &  \multirow{4}{2cm}{density, $\rho$, g/cm$^3$} & \multirow{4}{2.75cm}{Impacts number required for moderating, $n$} &  \multirow{4}{1.75cm}{Neutron mean free path, $\lambda$} &  \multirow{4}{1.75cm}{Moderator layer width, $R_{mod}$, cm} \\
    & & & & & &\\
    & & & & & &\\
    & & & & & &\\
    \hline    
    Be &   9 & 0.21 & 1.85 & 11 & 1.39 &15.3 \\
    \hline
    C   & 12 & 0.158 & 1.60 & 15 & 3.56 & 53.4 \\
    \hline
    H$_2$O & 18 & 0.924 & 1.0 & 2.5 & 16.7 & 41.6 \\
    \hline
    H$_2$O + B& & & & 2.5 &10.0 & 25.0 \\
    \hline
    He & 4 & 0.425 & 0.18 & 5.41 & 11.2 & 60.7 \\
    \hline
  \end{tabular}
\caption{Moderating and absorbing properties of some substances, moderator 
         layer width estimate for moderating neutron from $E_{fuel} = 1.0 ~MeV$ 
	 to $E_{mod} = 0.1 ~MeV$.}
\label{tab2}
\end{table}

And finally one can make the following intermediate conclusion. As shown above 
in Chapter~\ref{sec3}, the algorithm for determining the parameters 
(\ref{eq68a}) is mainly defined by para-parameter that plays a role of a 
``response function'' to all the physics of nuclear transformations, predefined 
by initial fuel composition. It is also very important that this parameter 
unequivocally determines the conditions of the nuclear burning wave existence 
(\ref{eq3a}), the neutron nuclear burning wave speed (see~(\ref{eq3a})) and 
the dimensionless width~(\ref{eq62}) of the supercritical area in the wave of
the active component burning. 

Based on the para-parameter ideology \cite{ref29} and Pavlovych group results
~\cite{ref20,ref21,ref26}, we managed to pick up a mode for the nuclear burning 
wave in U-Pu cycle, having the parameters shown in Table~\ref{tab1} satisfying
~(\ref{eq68a}). The latter means that the problem of dpa-parameter in cladding 
materials in the TWR-project is currently not an insurmountable technical 
problem and can be successfully solved.

In our opinion, the major problem of TWR are the so-called temperature blow-up 
modes that take place due to coolant loss as observed during Fukushima nuclear 
accident. Therefore below we shall consider the possible causes of the TWR 
inherent safety breach due to temperature blow-up mode.

\section{Possible causes of the TWR inherent safety failure: Fukushima plutonium
effect and the temperature blow-up mode}
\label{sec5}

It is known that with loss of coolant at three nuclear reactors during the 
Fukushima nuclear accident its nuclear fuel melted. It means that the 
temperature in the active zone reached the melting point of uranium-oxide fuel at some moments\footnote{Note that the third block partially used MOX-fuel 
enriched with plutonium}, i.e. $\sim$3000$^{\circ}$C. 

Surprisingly enough, in scientific literature today there are absolutely no 
either experimental or even theoretically calculated data on behavior of the 
$^{238}U$ and $^{239}Pu$ capture cross-sections depending on temperature at 
least in 1000-3000$^{\circ}$C range. At the same time there are serious reasons 
to suppose that the cross-section values of the specified elements increase with
temperature. We may at least point to qualitative estimates by Ukraintsev~
\cite{ref54}, Obninsk Institute of Atomic Energetics (Russia), that confirm the 
possibility of the cross-sections growth for $^{239}Pu$ in 300-1500$^{\circ}$C 
range.

Obviously, such anomalous temperature dependency of capture and fission 
cross-sections of $^{238}U$ and $^{239}Pu$ may change the neutron and thermal 
kinetics of a nuclear reactor drastically, including the perspective fast 
uranium-plutonium new generation reactors (reactors of Feoktistov~(\ref{eq1a}) 
and Teller~(\ref{eq2a}) type), which we classify as fast TWR reactors. Hence it is very important to know the anomalous temperature behavior of $^{238}U$ and 
$^{239}Pu$ capture and fission cross-sections, as well as their influence on the
heat transfer kinetics, because it may turn into a reason of the positive 
feedback\footnote{Positive Feedback is a type of feedback when a change in the 
output signal leads to such a change in the input signal, which leads to even 
greater deviation of the output signal from its original value. In other words, 
PF leads to the instability and appearance of qualitatively new (often 
self-oscilation) systems.} (PF) with the neutron kinetics leading to an 
undesirable solution stability loss (the nuclear burning wave), and consequently
to a trivial reactor runaway with a subsequent nontrivial catastrophe.

A special case of the PF is a non-linear PF, which leads to the system evolution
in the so-called blow-up mode \cite{ref55,ref56,ref57,ref58,ref59,ref60}, or in 
other words, in such a dynamic mode when one or several modeled values (e.g. 
temperature and neutron flux) grows to infinity at a finite time. In reality,
instead of the infinite values, a phase transition is observed in this case, 
which can become a first stage or a precursor of the future technogenic 
disaster.

Investigation of the temperature dependency of $^{238}U$ and $^{239}Pu$ capture 
and fission cross-sections in 300-3000$^{\circ}$C range and the corresponding 
kinetics of heat transfer and its influence on neutron kinetics in TWR is the 
main goal of the chapter.

Heat transfer equation for uranium-plutonium fissile medium is:

\begin{align}
\rho \left( \vec{r}, T, t \right) \cdot & c \left( \vec{r}, T, t \right) \cdot \dfrac{\partial T \left( \vec{r},t \right)}{\partial t} = \nonumber \\
& = \aleph \left( \vec{r},T,t \right) \cdot \Delta T \left( \vec{r},t \right) + \nabla \aleph \left( \vec{r},T,t \right) \cdot \nabla T \left( r,t \right) + q_T ^f \left( \vec{r},T,t \right),
\label{eq69}
\end{align}

\noindent where the effective substance density is

\begin{equation}
\rho \left( \vec{r},T,t \right) = \sum \limits_i N_i  \left( \vec{r},T,t \right) \cdot \rho_i,
\label{eq70}
\end{equation}

\noindent $\rho_i$ are tabulated values, $N_i \left( \vec{r},T,t \right)$ are 
the components concentrations in the medium, while the effective specific heat 
capacity (accounting for the medium components heat capacity values $c_i$) and 
fissile material heat conductivity coefficient (accounting for the medium 
components heat conductivity coefficients $\aleph_i (T)$) respectively are:

\begin{equation}
c \left( \vec{r},T,t \right) = \sum \limits_i c_i (T) N_i \left( \vec{r},T,t \right),
\label{eq71}
\end{equation}

\begin{equation}
\aleph \left( \vec{r},T,t \right) = \sum \limits_i \aleph_i (T) N_i \left( \vec{r},T,t \right).
\label{eq72}
\end{equation}

Here $q_T ^f \left( \vec{r},T,t \right)$ is the heat source density generated by
the nuclear fissions $N_i$ of fissile metal components that vary in time.

Theoretical temperature dependency of heat capacity $c(T)$ for metals is known:
at low temperatures $c(t) \sim T^3$, and at high temperatures $c(T) \rightarrow
const$, and the constant value ($const \approx 6 ~Cal/(mol \cdot deg)$) is 
determined by Dulong-Petit law. At the same time it is known that the thermal 
expansion coefficient is small for metals, therefore the specific heat capacity 
at constant volume $c_v$ almost equals to the specific heat capacity at constant
pressure $c_p$. On the other hand, the theoretical dependency of heat 
conductivity $\aleph_i (T)$ at high temperature of "fissile" metals is not 
known, while it is experimentally determined that the heat conductivity 
coefficient $\aleph (T)$ of fissile medium is a non-linear function of 
temperature (e.g. see~\cite{ref61}, where the heat conductivity coefficient is 
given for $\alpha$-uranium 238 and for metallic plutonium 239, and also~
\cite{ref62}).

While solving the heat conduction equations we used the following initial and 
boundary conditions:

\begin{equation}
T (r,t = 0) = 300~K ~~~and ~~ j_n = \aleph \left[ T(r \in \Re,t) - T_0 \right],
\label{eq73}
\end{equation}

\noindent where $j_n$ is the normal (to the fissile medium boundary) heat flux 
density component, $\aleph (T$) is the thermal conductivity coefficient, $\Re$ 
is the fissile medium boundary, $T_0$ is the temperature of the medium adjacent 
to the active zone.

Obviously, if the cross-sections of some fissile nuclides increase, then due to 
the nuclei fission reaction exothermicity, the direct consequence of 
the significantly non-linear kinetics of the parental and child nuclides in the 
nuclear reactor is an autocatalyst increase of generated heat, similar to 
autocatalyst processes of the exothermic chemical reactions. In this case the 
heat flux density $q_T ^f \left(\vec{r},\Phi,T,t \right)$ that characterizes the
generated heat amount will be:

\begin{equation}
q_T ^f \left(\vec{r},\Phi,T,t \right) = \Phi \left(\vec{r},T,t \right) \sum \limits_i Q_i^f \overline{\sigma}_f^i \left(\vec{r},T,t \right) N_i \left(\vec{r},T,t \right), ~~[W/cm^3],
\label{eq74}
\end{equation}

\noindent where

\begin{equation}
\Phi \left(\vec{r},T,t \right) = \int \limits_0 ^{E^{max}_n} \Phi \left(\vec{r},E,T,t \right) dE \nonumber
\end{equation}

\noindent is the full neutron flux density; $\Phi \left(\vec{r},E,T,t \right)$ 
is the neutron flux density with energy $E$; $Q_i ^f$ is the mean generated heat
emitted due to fission of one nucleus of the $i$-th nuclide;

\begin{equation}
 \overline{\sigma}_f^i \left(\vec{r},T,t \right) = \int \limits_0 ^{E_n^{max}} \sigma_f^i (E,T) \rho \left(\vec{r},E,T,t \right) dE  \nonumber
\end{equation}

\noindent is the fission cross-section of the $i$-th nuclide averaged over the 
neutron spectrum;

\begin{equation}
 \rho \left(\vec{r},E,T,t \right) = \Phi \left(\vec{r},E,T,t \right) / \Phi \left(\vec{r},T,t \right)  \nonumber
\end{equation}

\noindent is the neutron energy distribution probability density function; 
$\sigma_f^i (E,T)$ is the microscopic fission cross-section of the i-th nuclide 
that, as known, depends on the neutron energy and fissile medium temperature 
(Doppler effect~\cite{ref35}); $N_i \left(\vec{r},T,t \right)$ is the density of
the $i$-th nuclide nuclei.

As follows from (\ref{eq74}), in order to build the thermal source density 
function it is necessary to derive the theoretical dependency of the 
cross-sections $\overline{\sigma}_f^i \left(\vec{r},T,t \right)$, averaged over 
the neutron spectrum, on the reactor fuel temperature. As is known, the 
influence of the nuclei thermal motion on the medium comes to a broadening and 
height reduction of the resonances. By optical analogy, this phenomenon is 
referred to as Doppler effect~\cite{ref35}. Since the resonance levels are 
observed only for heavy nuclei in the low energy area, then Doppler effect is 
notable only during the interaction of neutrons with such nuclei. And the higher
environment temperature the stronger is the effect.

\begin{figure}
  \begin{center}
    \includegraphics[width=10cm]{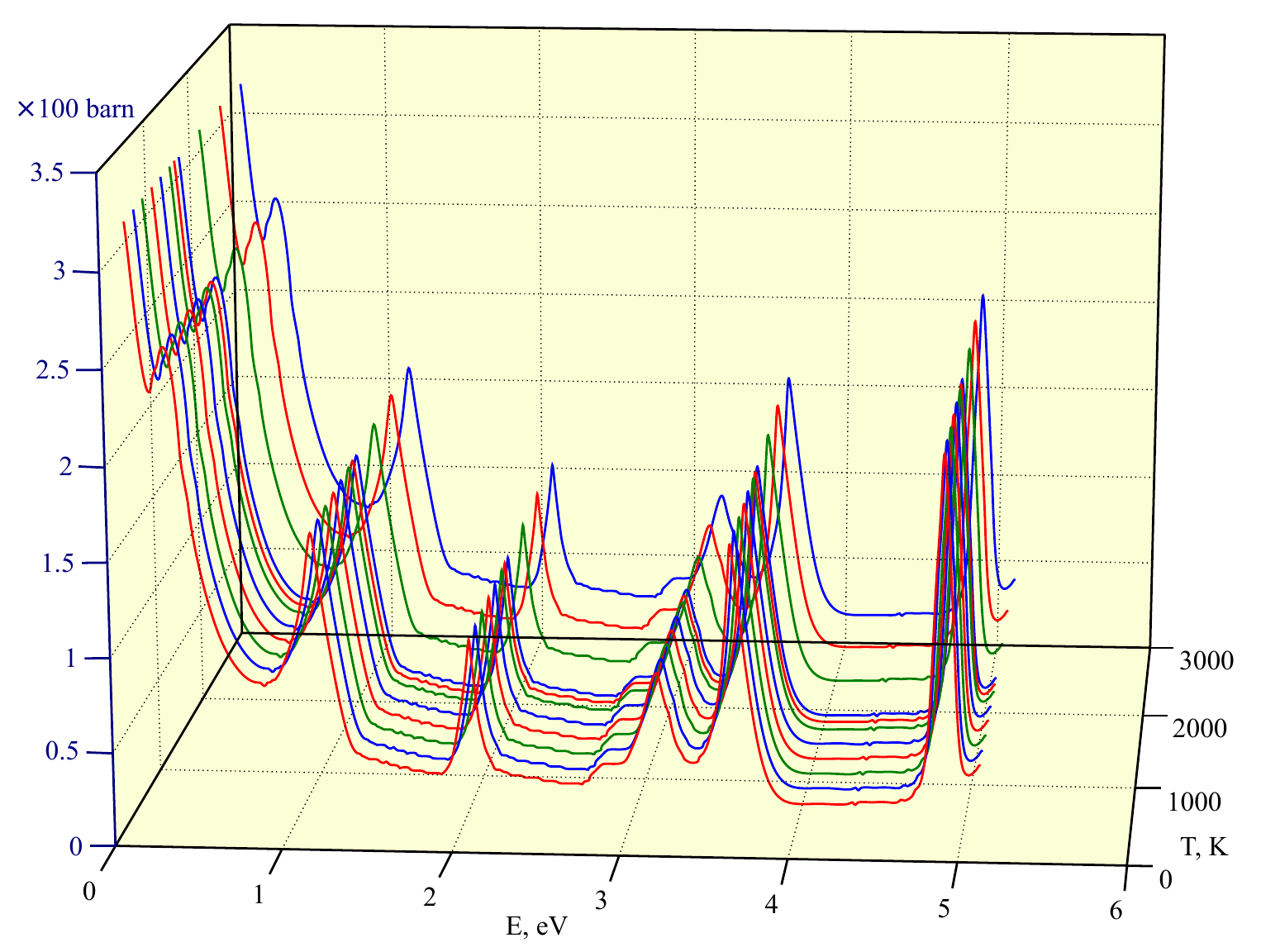}
    \caption{Calculated dependency of radioactive neutron capture cross-section
             on the energy for $^{235}_{92}U$ at different temperatures within 
	     300K to 3000K.}
    \label{fig11}
  \end{center}
\end{figure}

Therefore a program was developed using Microsoft Fortran Power Station 4.0 
(MFPS 4.0) that allows at the \underline{\textbf{first stage}} to calculate the 
cross-sections of the resonance neutron reactions depending on neutron energy 
taking into account the Doppler effect. The cross-sections dependency on neutron
energy for reactor nuclides from ENDF/B-VII database~\cite{ref63}, corresponding
to 300K environment temperature, were taken as the input data for the 
calculations. For example, the results for radioactive neutron capture 
cross-sections dependency on neutron energy for $^{235}U$ are given in 
fig.~\ref{fig11} for different temperatures of the fissile medium in 300K-3000K 
temperature range. Using this program, the dependency of scattering, fission and
radioactive neutron capture cross-sections for the major reactor fuel nuclides 
$_{92}^{235}U$, $_{92}^{238}U$, $_{92}^{239}U$ and $_{94}^239 Pu$  for different
temperatures in range 300K to 3000K were obtained.

At the \underline{\textbf{second stage}} a program was developed to obtain the 
calculated dependency of the cross-sections $\overline{\sigma}_f^i 
\left(\vec{r},T,t \right)$ averaged over the neutron spectrum for main reactor 
nuclides and for main neutron reactions for the specified temperatures. The 
averaging of the neutron cross-sections for the Maxwell distribution was 
performed using the following expression:

\begin{equation}
\left \langle \sigma \left( E_{lim},T \right) \right \rangle = \dfrac{\int \limits_0 ^{E_{lim}} E^{1/2} e^{-E/kT} \sigma (E,T) dE}{\int \limits_0 ^{E_{lim}} E^{1/2} e^{-E/kT} dE}, \nonumber
\end{equation}

\noindent where $E_{lim}$ is the upper limit of the neutrons thermalization, 
while for the procedure of neutron cross-sections averaging over the Fermi 
spectrum the following expression was used: 

\begin{equation}
\left \langle \sigma \left( E_{lim},T \right) \right \rangle = \dfrac{\int \limits_{E_{lim}}^{\infty} \sigma (E,T) E^{-1} dE}{\int \limits_{E_{lim}}^{\infty} E^{-1} dE}, \nonumber
\end{equation}

During further calculations in our programs we used the results obtained at the 
first stage i.e. the dependency of reaction cross-sections on neutron energy and
environment temperature (Doppler effect). The neutron spectrum was specified in 
a combined way -- by Maxwell spectrum $\Phi_M \left( E_n \right)$ below the 
limit of thermalization $E_{lim}$; by Fermi spectrum $\Phi_F (E)$ for a 
moderating medium with absorption above $E_{lim}$ but below $E_F$ (upper limit 
for Fermi neutron energy spectrum); by $^{239}Pu$ fission spectrum \cite{ref21,
ref22} above $E_F$, but below the maximal neutron energy $E_n^{max}$. Here the 
neutron gas temperature for Maxwell distribution was given by (\ref{eq75}), 
described in \cite{ref35}. According to this approach~\cite{ref35}, the 
drawbacks of the standard slowing-down theory for thermalization area may be 
formally reduced if a variable $\xi (x) = \xi (1 - 2/z)$ is introduced instead 
of the average logarithmic energy loss $\xi$, which is almost independent of the
neutron energy (as is known, the statement $\xi \approx 2/A$ is true for the 
environment consisting of nuclei with $A > 10$). Here $z = E_n / kT$, $E_n$ is 
the neutron energy, $T$ is the environment temperature. Then the following 
expression may be used for the neutron gas temperature in Maxwell spectrum of 
thermal neutrons\footnote{A very interesting expression revealing hidden 
connection between the temperature of a neutron gas and the environment (fuel) 
temperature.}:

\begin{equation}
T_n = T_0 \left[ 1 + \eta \cdot \dfrac{\Sigma_a (k T_0)}{\langle \xi \rangle \Sigma_S} \right],
\label{eq75}
\end{equation}

\noindent where $T_0$ is the fuel environment temperature, $\Sigma_a (k T_0)$ is
an absorption cross-section for energy $k T_0$, $\eta = 1.8$ is the 
dimensionless constant, $\langle \xi \rangle$ is averaged over the whole energy 
interval of Maxwell spectrum $\xi (z)$ at $kT = 1~eV$.

Fermi neutron spectrum for a moderating medium with absorption (we considered 
carbon as a moderator and $^{238}U$, $^{239}U$ and $^{239}Pu$ as the absorbers) 
was set in the form \cite{ref35,ref64}:

\begin{equation}
\Phi_{Fermi} \left(E, E_F \right) = \dfrac{S}{\langle \xi \rangle \Sigma_t E} \exp {\left[ - \int \limits_{E_{lim}}^{E_f} \dfrac{\Sigma_a \left( E' \right) dE'}{\langle \xi \rangle \Sigma_t \left(E' \right) E'} \right]},
\label{eq76}
\end{equation}

\noindent where $S$ is the total volume neutron generation rate, $\langle \xi 
\rangle = \sum \limits_i \left( \xi_i \Sigma_S^i \right) / \Sigma_S$, $\xi_i$ is
the average logarithmic decrement of energy loss, $\Sigma_S^i$ is the 
macroscopic scattering cross-section of the $i$-th nuclide, $\Sigma_t = \sum 
\limits_i \Sigma_S^i + \Sigma_a^i$ is the total macroscopic cross-section of the
fissile material, $\Sigma_S = \sum \limits_i \Sigma_S^i$ is the total 
macroscopic scattering cross-section of the fissile material, $\Sigma_a$ is the 
macroscopic absorption cross-section, $E_F$ is the upper neutron energy for 
Fermi spectrum.

The upper limit of neutron thermalization $E_{lim}$ in our calculation was 
considered a free parameter, setting the neutron fluxes of Maxwell and Fermi 
spectra at a common energy limit $E_{lim}$ equal:

\begin{equation}
\Phi_{Maxwell} \left( E_{lim} \right) = \Phi_{Fermi} \left( E_{lim} \right).
\label{eq77}
\end{equation}

The high energy neutron spectrum part ($E > E_F$) was defined by the fission 
spectrum \cite{ref64,ref65,ref66} in our calculations. Therefore the following 
expression may be written for the total volume neutron generation rate $S$ in 
the Fermi spectrum (\ref{eq76}):

\begin{equation}
S \left( \vec{r},T,t \right) = \int \limits_{E_F}^{E_n^{max}} \tilde{P} \left( \vec{r},E,T,t \right) \left[ \sum \limits_i \nu_i (E) \cdot \Phi \left( \vec{r},E,T,t \right) \cdot \sigma_f^i (E,T) \cdot N_i \left(\vec{r},T,t \right) \right] dE,
\label{eq78}
\end{equation}

\noindent where $E_n^{max}$ is the maximum energy of the neutron fission 
spectrum (usually taken as $E_n^{max} \approx 10~MeV$), $E_F$ is the neutron 
energy, below which the moderating neutrons spectrum is described as Fermi 
spectrum (usually taken as $E_F \approx 0.2~MeV$); $\tilde{P} \left( 
\vec{r},E,T,t \right)$ is the probability of neutron not leaving the boundaries 
of the fissile medium which depends on the fissile material geometry and the
conditions at its border (e.g. presence of a reflector).

\begin{figure}
  \begin{center}
    \begin{minipage}[c]{0.45\linewidth}
      \begin{center}
      \includegraphics[width=7cm]{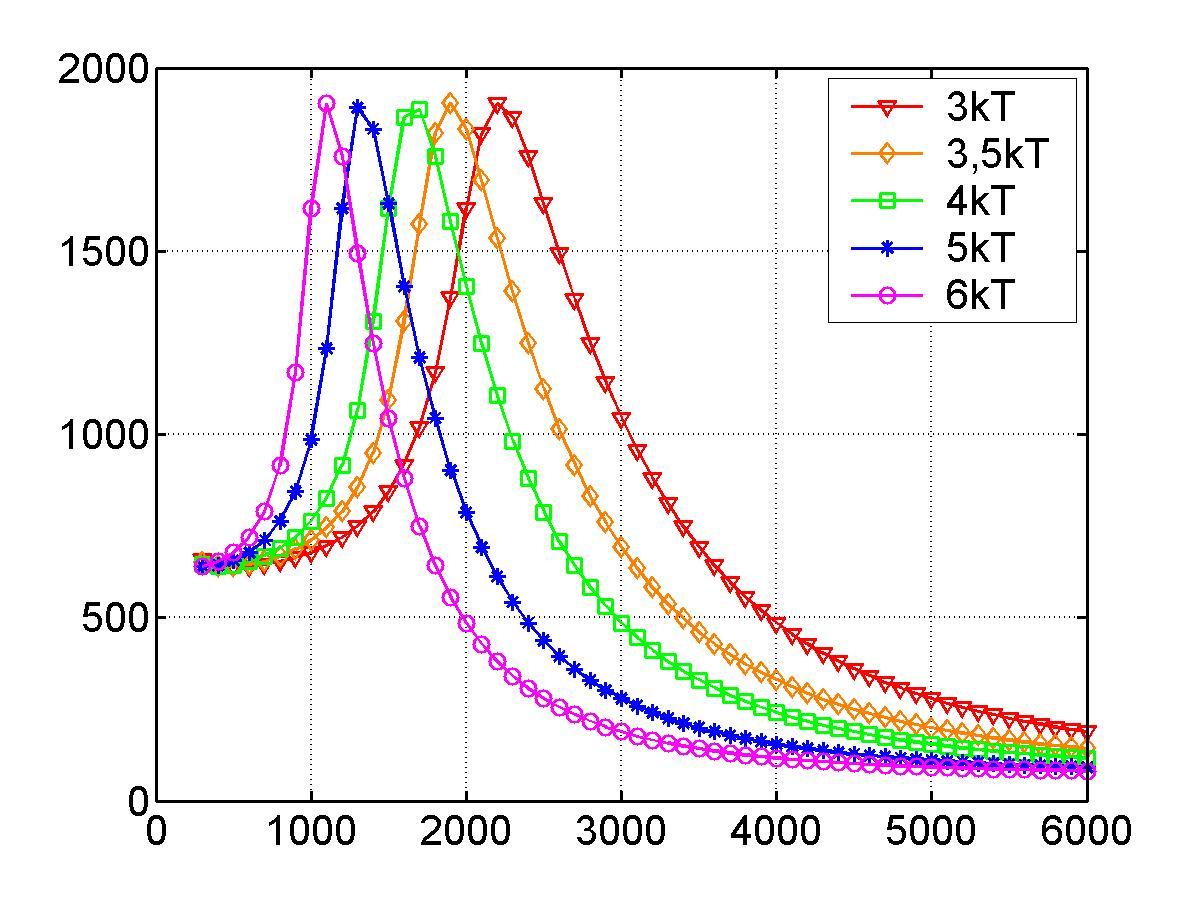} \\ a)
      \end{center}
    \end{minipage}
    \hfill
    \begin{minipage}[c]{0.45\linewidth}
      \begin{center}
      \includegraphics[width=7cm]{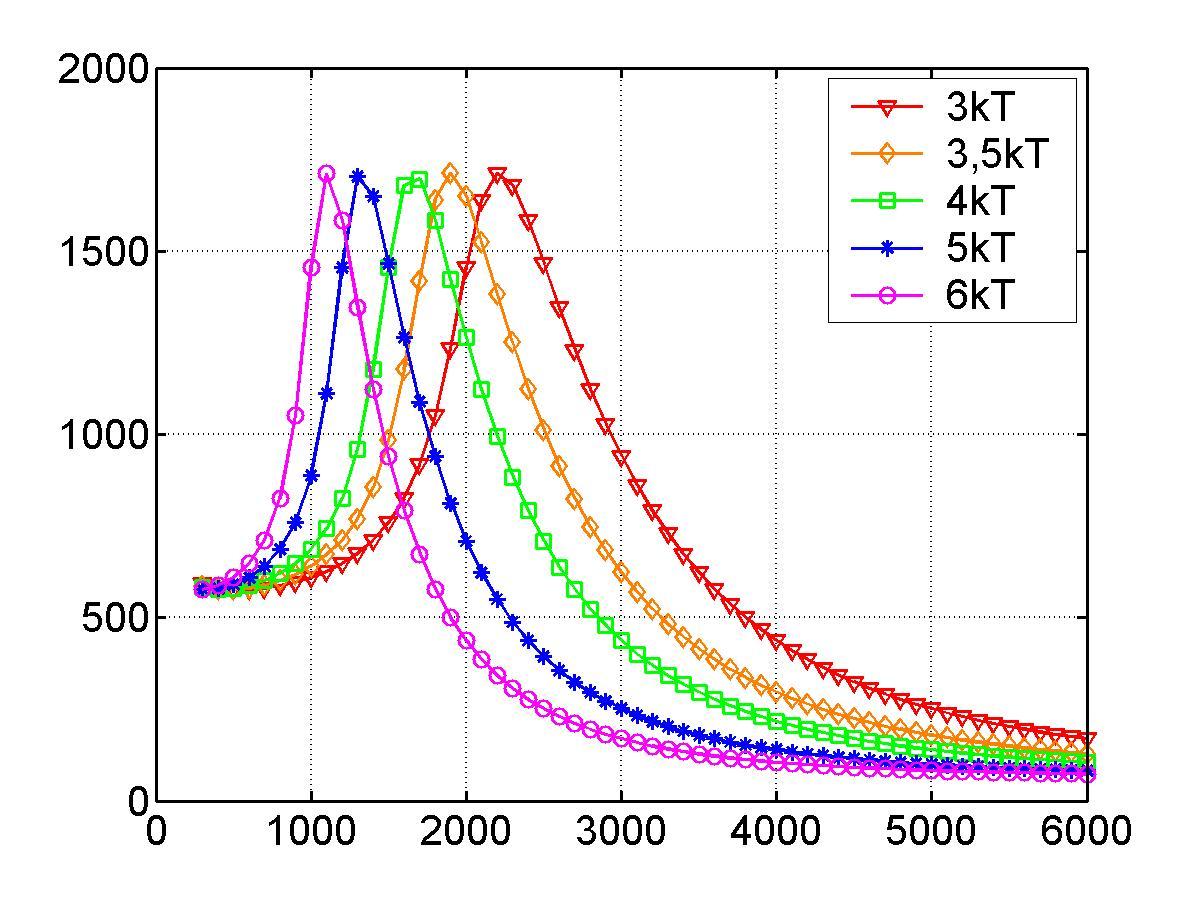} \\ b)
      \end{center}
    \end{minipage}
    \caption{Temperature dependencies for the fission cross-section (a) and 
             radioactive capture cross-section (b) for $^{239}Pu$, averaged over
	     the Maxwell spectrum, on the Maxwell and Fermi spectra joining 
	     energy and $\eta = 1.8$ (see (\ref{eq75})).}
    \label{fig12}
  \end{center}
\end{figure}

\begin{figure}
  \begin{center}
        \begin{minipage}[c]{0.45\linewidth}
      \begin{center}
      \includegraphics[width=7cm]{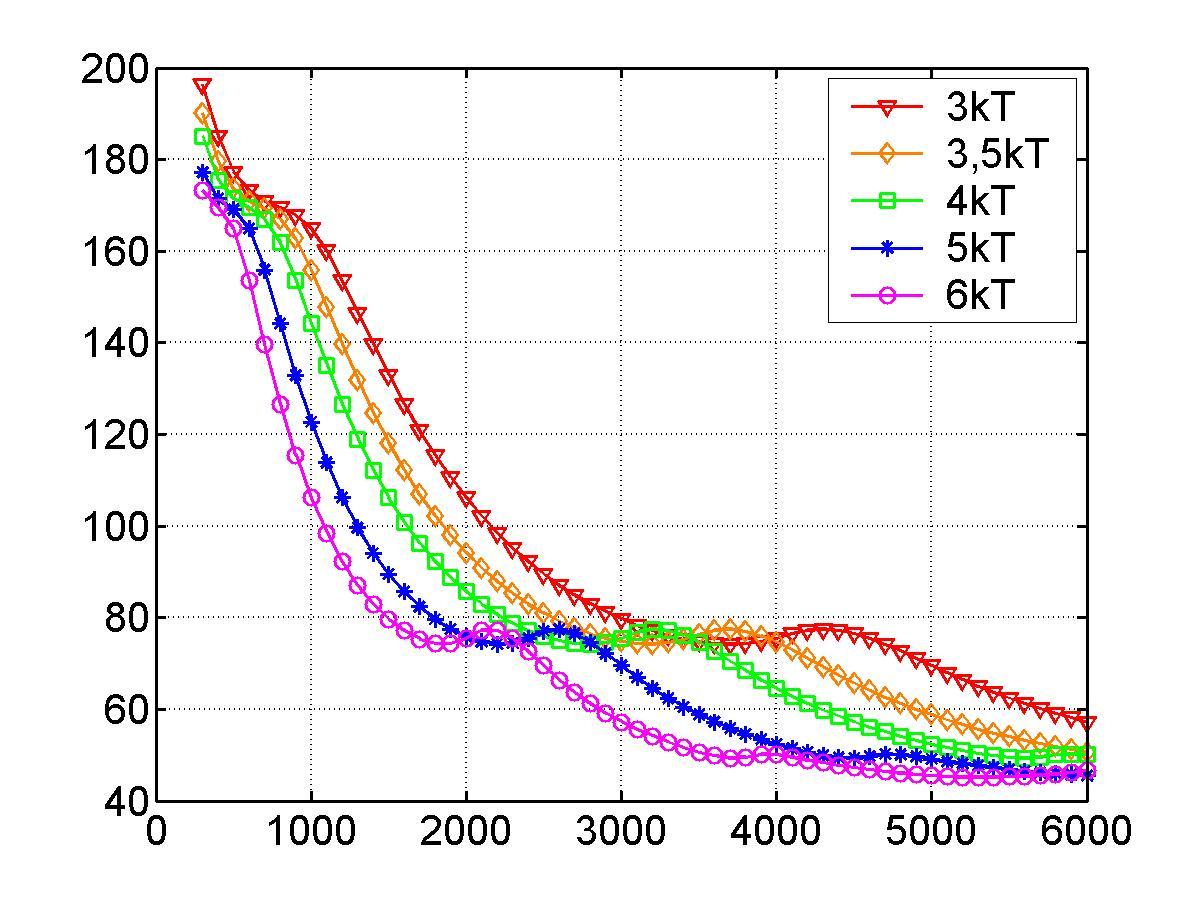} \\ a)
      \end{center}
    \end{minipage}
    \hfill
    \begin{minipage}[c]{0.45\linewidth}
      \begin{center}
      \includegraphics[width=7cm]{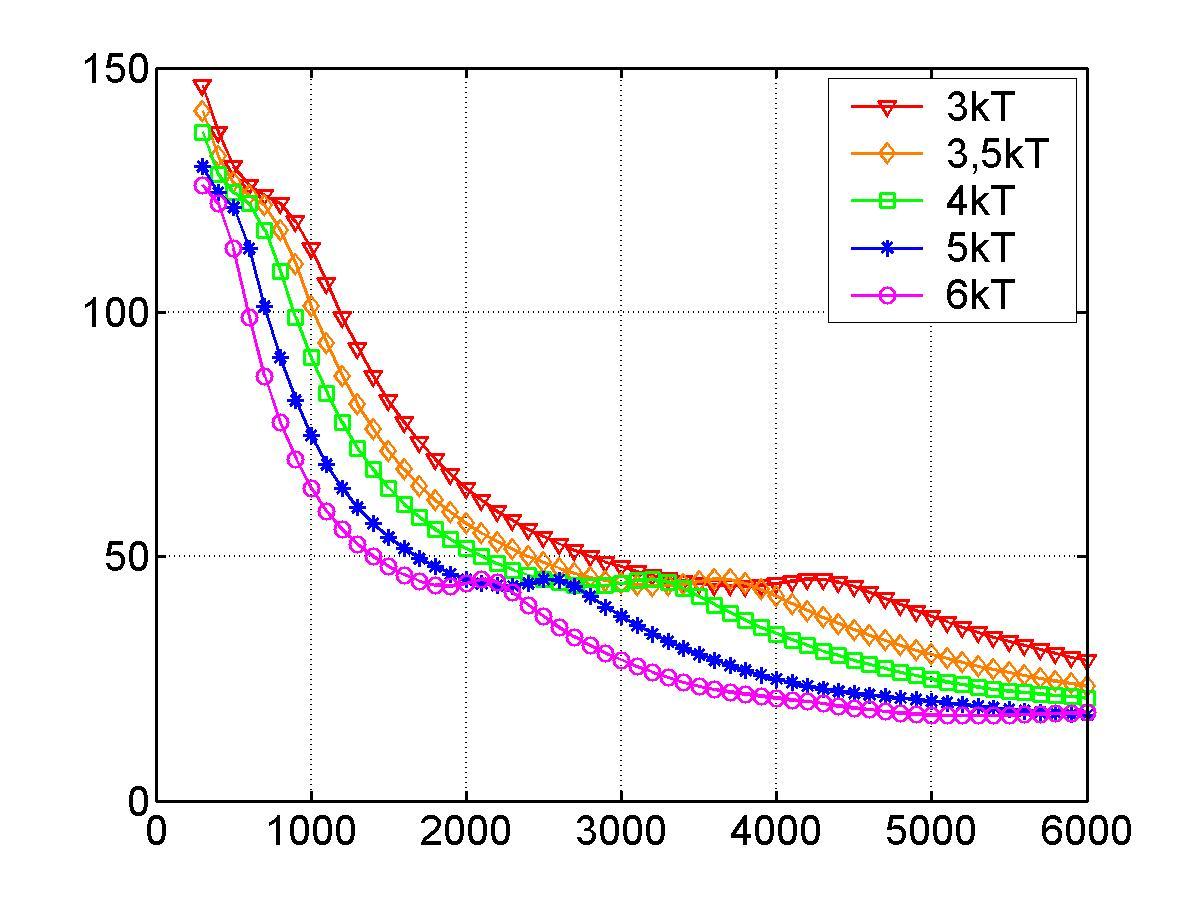} \\ b)
      \end{center}
    \end{minipage}
    \caption{Temperature dependencies for the fission cross-section (a) and 
             radioactive capture cross-section (b) for $^{235}U$, averaged over 
	     the Maxwell spectrum, on the Maxwell and Fermi spectra joining 
	     energy and $\eta = 1.8$ (see (\ref{eq75})).}
    \label{fig13}
  \end{center}
\end{figure}

\begin{figure}
  \begin{center}
        \begin{minipage}[c]{0.45\linewidth}
      \begin{center}
      \includegraphics[width=7cm]{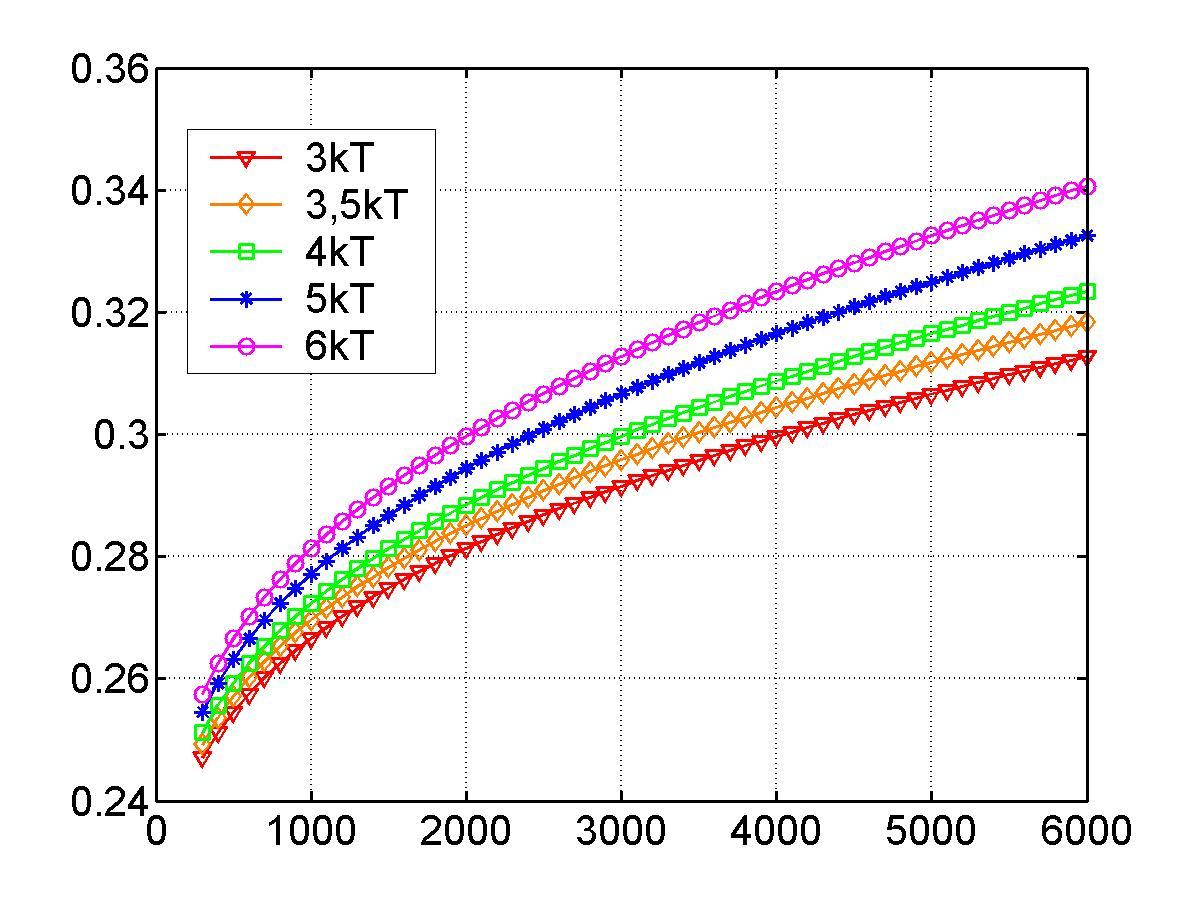} \\ a)
      \end{center}
    \end{minipage}
    \hfill
    \begin{minipage}[c]{0.45\linewidth}
      \begin{center}
      \includegraphics[width=7cm]{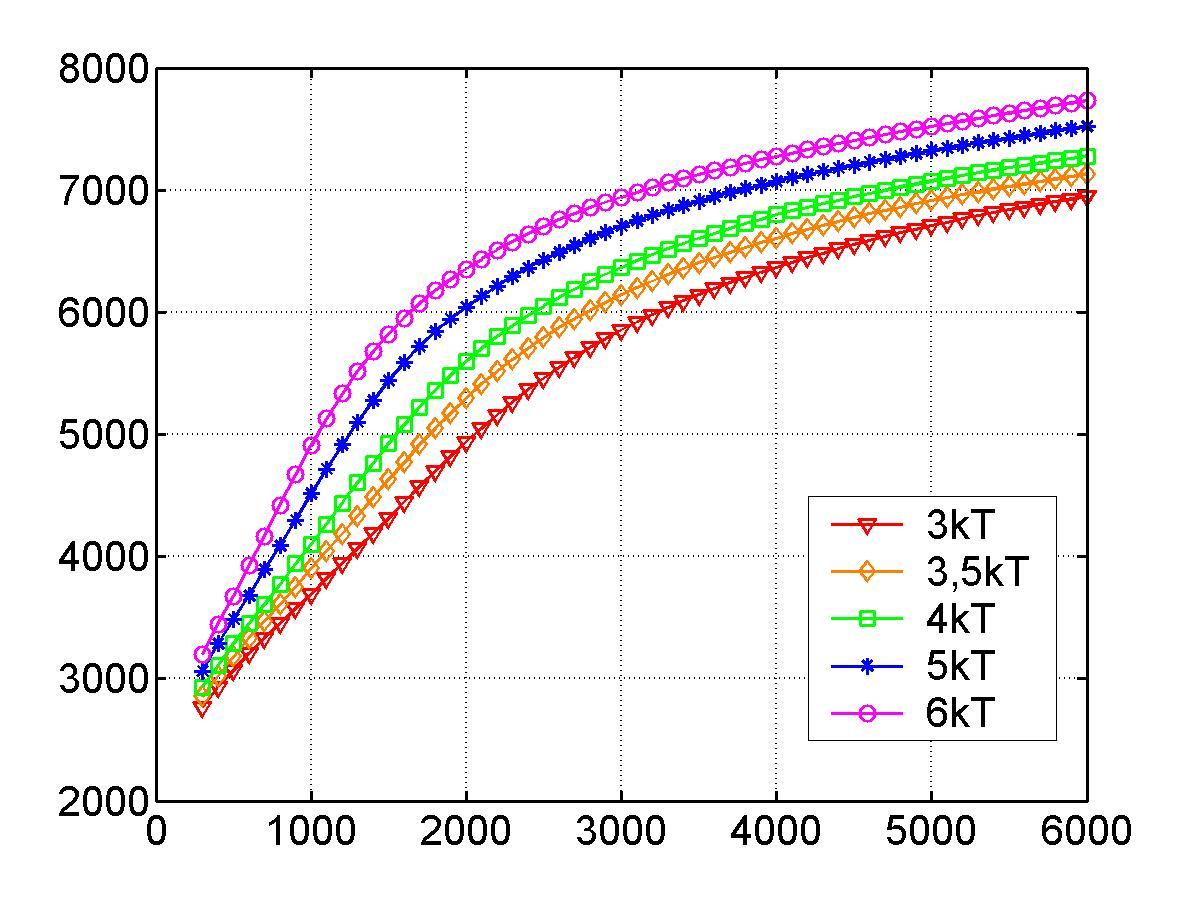} \\ b)
      \end{center}
    \end{minipage}
    \caption{Temperature dependencies for the fission cross-section (a) and 
             radioactive capture cross-section (b) for $^{238}_{92}U$, averaged 
	     over the combined Maxwell and Fermi spectra depending on the 
	     Maxwell and Fermi spectra joining energy and $\eta = 1.8$ (see 
	     (\ref{eq75})).}
    \label{fig14}
  \end{center}
\end{figure}

The obtained calculation results show that the cross-sections averaged over the 
spectrum may increase (fig.~\ref{fig12} for $^{239}Pu$ and fig.~\ref{fig14} for 
$^{238}U$) as well as decrease (fig.~\ref{fig13} for $^{235}U$) with fissile 
medium temperature. As follows from the obtained results, the arbitrariness in 
selection of the limit energy for joining the Maxwell and Fermi spectra does not
significantly alter the character of these dependencies evolution.

This can be justified by the fact that $^{239}Pu$ resonance area starts from 
significantly lower energies than that of $^{235}U$, and with fuel temperature 
increase the neutron gas temperature also increases producing the Maxwell's 
neutron distribution maximum shift to the higher neutron energies. I.e. the 
neutron gas spectrum hardening, when more neutrons fit into resonance area of 
$^{239}Pu$, is the cause of the averaged cross-sections growth.

This process in not as significant for $^{235}U$ because its resonance area is 
located at higher energies. As a result, the $^{235}U$ neutron gas spectrum 
hardening related to the fuel temperature increase (in the considered interval) 
does not result in a significant increase of a number of neutrons fitting into 
the resonance area. Therefore according to the known expressions for $^{235}U$ 
determining the neutron reactions cross-sections behaviour depending on their 
energy $E_n$ for non-resonance areas, we observe dependency for the averaged 
cross-sections $\sigma_{nb} \sim 1/\sqrt{E_n}$.

The data on the averaged fission and capture cross-sections of $^{238}U$ 
presented at fig.~\ref{fig14} show that the averaged fission cross-section for 
$^{238}U$ is almost insensitive to the neutron spectrum hardening caused by the 
fuel temperature increase -- due to a high fission threshold $\sim$1~MeV (see 
fig.~\ref{fig14}a). At the same time they confirm the capture cross-section 
dependence on temperature, since its resonance area is located as low as for 
$^{239}Pu$. Obviously, in this case the fuel enrichment with $^{235}U$ makes no 
difference because the averaged cross-sections for $^{235}U$, as described 
above, behave in a standard way.

\begin{figure}
  \begin{center}
    \includegraphics[width=10cm]{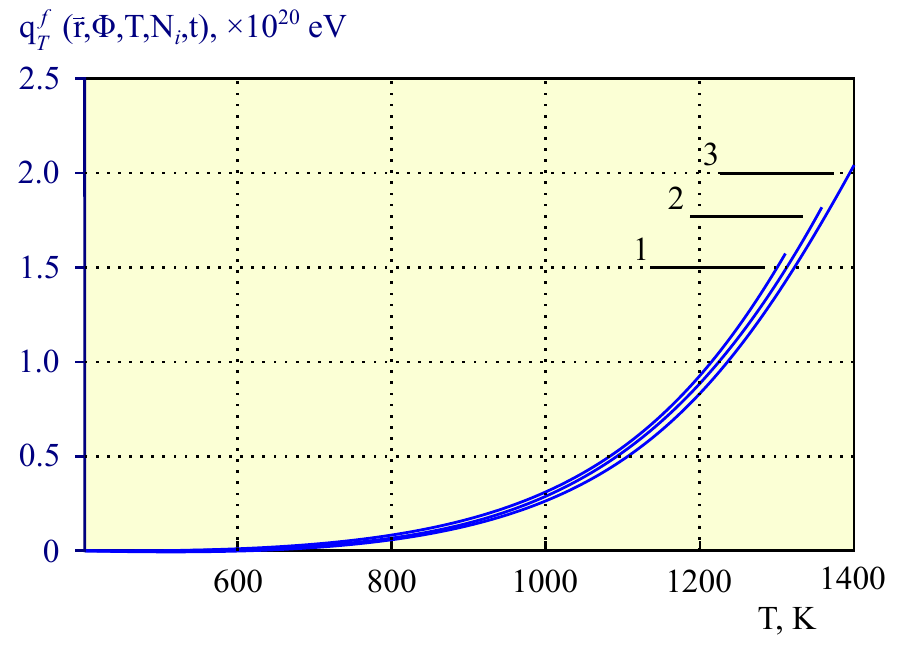}
    \caption{Dependence of the heat source density  $q_T ^f \left( r, \Phi, T, 
             N_i, t \right)~[eV]$ on the fissile medium temperature (300-1400K) 
	     for several compositions of uranium-plutonium medium (1 - 10\%~Pu;
	     2 - 5\%~Pu; 3 - 1\%~Pu) at the constant neutron flux density 
	     $\Phi = 10^{13}~n/(cm^2 \cdot s)$.}
    \label{fig15}
  \end{center}
\end{figure}

And finally we performed a computer estimate of the heat source density 
dependence $q_T^f \left(\vec{r},\Phi,T,t \right)$ (\ref{eq74}) on temperature 
for the different compositions of the uranium-plutonium fissile medium with a 
constant neutron flux density, presented at fig.~\ref{fig15}. We used the 
dependencies presented above at fig.\ref{fig12}-\ref{fig14} for these 
calculations. Let us note that our preliminary calculations were made not taking
into account the change in the composition and density of the fissile 
uranium-plutonium medium, which is a direct consequence of the constant neutron 
flux assumption.

The necessity of such assumption is caused by the following. The reasonable 
description of the heat source density $q_T^f \left(\vec{r},\Phi,T,t \right)$ 
(\ref{eq74}) temperature dependence requires the solution of a system of three 
equations -- two of them correspond to the neutron kinetics equation (flux and 
fluence) and to the system of equations for kinetics of the parental and child 
nuclides nuclear density (e.g. see~\cite{ref29,ref26}), while the third one 
corresponds to a heat transfer equation of (\ref{eq69}) type. However, some 
serious difficulties arise here associated with the computational capabilities 
available. And here is why.

One of the principal physical peculiarities of the TWR is the fact~\cite{ref20} 
that fluctuation residuals of plutonium (or $^{233}U$ in Th-U cycle) over its 
critical concentration burn out for the time comparable with reactor lifetime of
a neutron $\tau_n (x,t)$ (not considering the delayed neutrons), or at least 
comparable with the reactor period\footnote{The reactor period by definition 
equals to $T(x,t) = \tau_n (x,t) / \rho(x,t)$, i.e. is a ratio of the reactor 
neutron lifetime to reactivity.} $T(x,t)$ (considering the delayed neutrons). 
Meanwhile the new plutonium (or $^{233}U$ in Th-U cycle) is formed in a few days
(or a month) and not immediately. This means \cite{ref20} that the numerical 
calculation must be made with a temporal step about 10$^{-6}$-10$^{-7}$ in case 
of not taking into account the delayed neutrons and $\sim$10$^{-1}$-10$^{0}$ 
otherwise. At first glance, taking into account the delayed neutrons, according 
to \cite{ref20}, really ``saves the day'', however it is not always true. If the
heat transfer equation contains a significantly non-linear source, then in the 
case of a blow-up mode, the temperature may grow extremely fast under some 
conditions and in 10-20 steps (with time step 10$^{-6}$-10$^{-7}$ s) reaches 
the critical amplitude that may lead to (at least) a solution stability loss, or
(as maximum) to a blow-up bifurcation of the phase state, almost unnoticeable 
with a rough time step.


According to these remarks, and considering the goal and format of this paper, 
we did not aim at finding the exact solution of some specific system of three 
joint equations described above. Instead, we found it important to illustrate --
at the qualitative level -- the consequences of the possible blow-up modes in 
case of a non-linear heat source presence in the heat transfer equation. As said
above, we made some estimate computer calculations of the heat source density 
$q_T^f \left(\vec{r},\Phi,T,t \right)$~(\ref{eq74}) temperature dependence in 
300-1400K range for some compositions of uranium-plutonium fissile medium at a
constant neutron flux (fig.~\ref{fig15}).

The obtained dependencies for the heat source density $q_T^f \left(\vec{r},\Phi,
T,t \right)$ were successfully approximated by a power function of temperature 
with an exponent of 4 (fig.~\ref{fig15}). In other words, we obtained a heat
transfer equation with a significantly nonlinear heat source in the following 
form:

\begin{equation}
q_T (T) = const \cdot T^{(1 + \delta)},
\label{eq79}
\end{equation}

\noindent where $\delta > 1$ in case of non-linear thermal conductivity 
dependence on temperature \cite{ref55,ref56,ref57,ref58,ref59}. The latter means
that the solutions of the heat transfer equation (\ref{eq69}) describe the 
so-called Kurdyumov blow-up modes \cite{ref55,ref56,ref57,ref58,ref59,ref60}, 
i.e. such dynamic modes when one of the modeled values (e.g. temperature) turns 
into  infinity for a finite time. As noted before, in reality instead of 
reaching the infinite values, a phase transition is observed (a final phase of 
the parabolic temperature growth), which requires a separate model and is a 
basis for an entirely new problem.

Mathematical modeling of the blow-up modes was performed mainly using 
Mathematica 5.2-6.0, Maple 10, Matlab 7.0, utilizing multiprocessor calculations
for effective application. A Runge–Kutta method of 8-9$^{th}$ order and the 
numerical methods of lines~\cite{ref67} were applied for the calculations. The 
numerical error estimate did not exceed 0.01\%. The coordinate and temporal 
steps were variable and chosen by the program in order to fit the given error at
every step of the calculation.

Below we give the solutions for the heat transfer equation (\ref{eq69}) with 
nonlinear exponential heat source (\ref{eq79}) in uranium-plutonium fissile 
medium for boundary and initial parameters corresponding to the industrial 
reactors. The calculations were done for a cube of the fissile material with 
different sizes, boundary and initial temperature values. Since the temperature 
dependencies of the heat source density were obtained without account for the
changing composition and density of the uranium-plutonium fissile medium, 
different blow-up modes can take place (HS-mode, S-mode, LS-mode) depending on 
the ratio between the exponents of the heat conductivity and heat source 
temperature dependences, according to \cite{ref55,ref56,ref57,ref58,ref59,
ref60}. Therefore we considered the cases for 1$^{\text{st}}$, 2$^{\text{nd}}$ 
and 4$^{\text{th}}$ temperature order sources. Here the power of the source also
varied by varying the proportionality factor in (\ref{eq79}) ($const = 1.00 J / 
(cm^3 \cdot s \cdot K$) for the 1$^{\text{st}}$ temperature order source; 
$0.10 J / (cm^3 \cdot s \cdot K^2)$, $0.15~J / (cm^3 \cdot s \cdot K^2)$ and 
$1.00~J / (cm^3 \cdot s \cdot K^2)$ for the 2$^{\text{nd}}$ temperature order 
source; $1.00~J / (cm^3 \cdot s \cdot K^4)$ for the 4$^{\text{th}}$ temperature 
order source).

\begin{figure}
  \begin{center}
    \includegraphics[width=10cm]{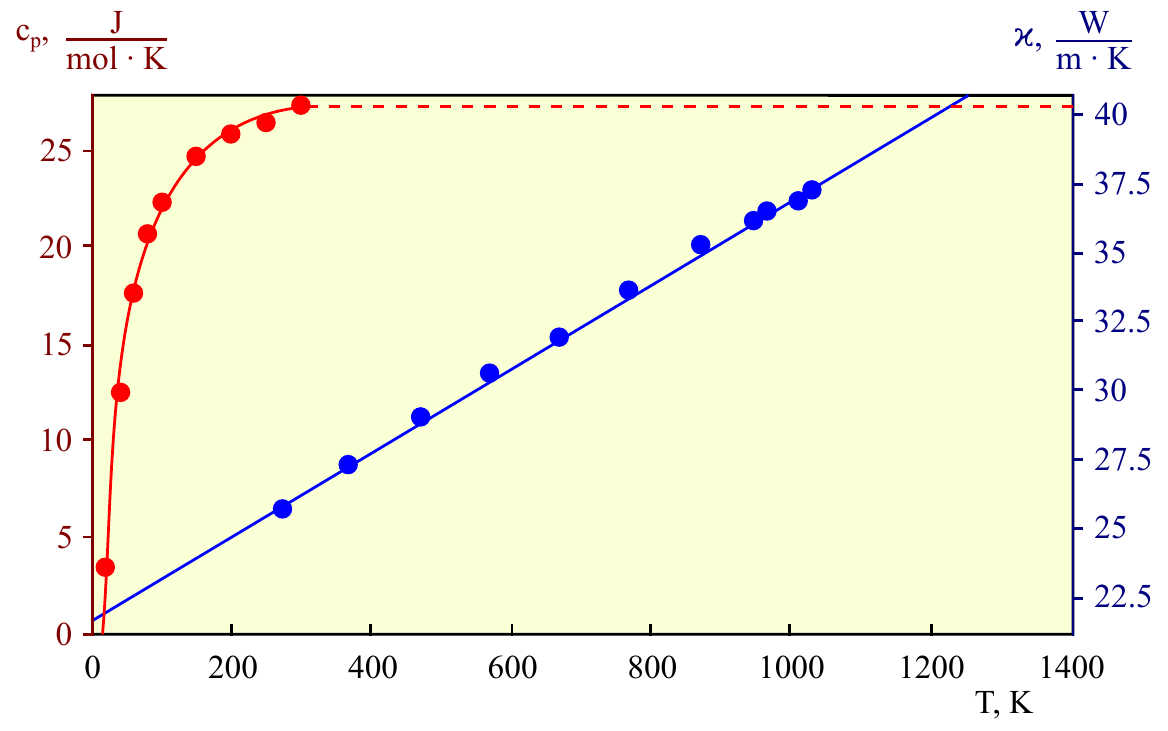}
    \caption{Temperature dependence of the heat capacity $c_P$ and heat 
             conductivity $\chi$ of the fissile material. Points represent the
	     experimental values for the heat capacity and heat conductivity of 
	     $^{238}U$.}
    \label{fig16}
  \end{center}
\end{figure}

During the calculations of the heat capacity $c_p$ (fig.~\ref{fig16}a) and heat 
conductivity $\aleph$ (fig.~\ref{fig16}b) of a fissile medium dependence on 
temperature in 300-1400K range the specified parameters were given by analytic 
expressions, obtained by approximation of experimental data for$^{238}U$ based 
on polynomial progression:

\begin{equation}
c_p (T) \approx - 7.206 + 0.64 T - 0.0047 T^2 + 0.0000126 T^3 + 2.004 \cdot 10^{-8} T^4 - 1.60 \cdot 10^{-10} T^5 - 2.15 \cdot 10^{-13} T^6,
\label{eq80}
\end{equation}

\begin{equation}
\aleph (T) \approx 21.575 + 0.0152661 T.
\label{eq81}
\end{equation}

\begin{figure}
  \begin{center}
    \includegraphics[width=17cm]{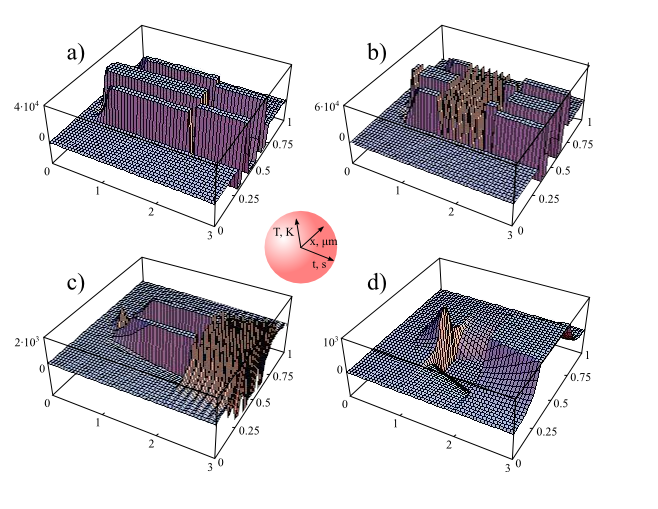}
    \caption{Heat transfer equation (\ref{eq69}) solution for 3D case (crystal 
             sizes 0.001$\times$0.001$\times$0.001~mm; initial and boundary 
	     temperatures equal to 100K): a) The source is proportional to the 
	     4$^{\text{th}}$ order of temperature; const = 1.00~$J / (cm^3 
	     \cdot s \cdot K^4)$, heat capacity and heat conductivity are 
	     constant and equal to 11.5~$ J / (K \cdot mol)$ and 27.5~$W / (m 
	     \cdot K)$ respectively; b) the source is proportional to the 
	     4$^{\text{th}}$ order of temperature; const = 1.00~$J / (cm^3 
	     \cdot s \cdot K^4)$; c) The source is proportional to the 
	     2$^{\text{nd}}$ order of temperature; const = 1.00~$J / (cm^3 
	     \cdot s \cdot K^2)$; d) the source is proportional to the 
	     2$^{\text{nd}}$ order of temperature; const = 0.10~$J / (cm^3 
	     \cdot s \cdot K^2)$. Note: in the cases b) - d) the heat capacity 
	     and heat conductivity were determined by (\ref{eq80}) and 
	     (\ref{eq81}) respectively.}
    \label{fig17}
  \end{center}
\end{figure}

And finally the heat transfer equation (\ref{eq69}) solution was obtained for 
the constant heat conductivity ($27.5~W / (m \cdot K)$) and heat capacity 
($11.5~J / (K \cdot mol)$), presented in fig.~\ref{fig17}a, and also the 
solutions of the heat transfer equation considering their temperature 
dependencies (fig.~\ref{fig17}b-d).

These results point directly to a possibility of the local uranium-plutonium 
fissile medium melting, with the melting temperature almost identical to that 
of $^{238}U$, which is 1400K (fig.~\ref{fig16}a-d). Moreover, these regions of 
the local melting are not the areas of the so-called thermal peaks~\cite{ref68},
and probably are the anomalous areas of uranium surface melting observed by 
Laptev and Ershler~\cite{ref69} that were also mentioned in~\cite{ref70}. More 
detailed analysis of the probable temperature scenarios associated with the 
blow-up modes are discussed below.

\section{The blow-up modes in neutron-multiplying media and the pulse 
thermonuclear TWR.}
\label{sec6}

Earlier we noted the fact that due to a coolant loss at the nuclear reactors 
during the Fukushima nuclear accident the fuel was melted, which means that the 
temperature inside the active zone reached the melting temperature of 
uranium-oxide fuel at some moment, i.e. $\sim$3000$^{\circ}$C. 

On the other hand, we already know that the coolant loss may become a cause of 
the nonlinear heat source formation inside the nuclear fuel, and therefore 
become a cause of the temperature and neutron flux blow-up mode onset. A natural
question arises of whether it is possible to use such blow-up mode (temperature 
and neutron flux) for the initiation of certain controlled physical conditions 
under which the nuclear burning wave would regularly ``experience'' the 
so-called ``controlled blow-up'' mode. It is quite difficult to answer this 
question definitely, because such fast process has a number of important 
physical vaguenesses, any of which can become experimentally insurmountable for 
such process control.

Nevertheless such process is very elegant and beautiful from the physics point 
of view, and therefore requires a more detailed phenomenological description. 
Let us try to make it in short.

As we can see from the plots of the capture and fission cross-sections 
evolution for $^{239}Pu$ (fig.~\ref{fig12}), the blow-up mode may develop 
rapidly at $\sim$1000-2000K (depending on the real value of the Fermi and 
Maxwell spectra joining boundary), but at the temperatures over 2500-3000K 
the cross-sections return almost to the initial values. If some effective heat
sink is turned on at that point, the fuel may return to its initial temperature.
However, while the blow-up mode develops, the fast neutrons already penetrate to
the adjacent fuel areas, where the new fissile material starts accumulating and 
so on (see cycles (\ref{eq1a}) and (\ref{eq2a})). After some time the similar
blow-up mode starts developing in this adjacent area and everything starts over 
again. In other words, such hysteresis blow-up mode, closely time-conjugated to 
a heat takeoff procedure, will appear on the background of a stationary nuclear 
burning wave in a form of the periodic impulse bursts.

In order to demonstrate the marvelous power of such process, we investigated the
heat transfer equation with non-linear exponential heat source in 
uranium-plutonium fissile medium with boundary and initial parameters emulating 
the heat takeoff process. In other words, we investigated the blow-up modes in 
the Feoktistov-type uranium-plutonium reactor (\ref{eq1a}), where the 
temperature inside and at the boundary was deliberately fixed at 6000K, which
corresponds to the model of the georeactor\footnote{Let us note that our model
georeactor is not a fast reactor. The possibility of the nuclear wave burning
for a reactor another than the fast one is examined in our next paper~\cite{ref76}}
\cite{ref17}. Expression (\ref{eq75})
for the neutron gas temperature, used for the calculation of the cross-sections 
averaged over the neutron spectrum, transforms in this case to the following:

\begin{equation}
T_n \approx \left[ 1 + 1.8 \frac{8.0 \cdot K_2}{<\xi> \cdot 4.5} \right]
\label{eq83}
\end{equation}

This equation is obtained for the supposed fissile medium composition of the
Uranium and Plutonium dicarbides~\cite{ref10a,ref10b,ref10c,ref17,ref75}, where
the $^{238}$U was the major absorber (its microscopic absorption cross-section
for the thermalization temperatures was set at $\sigma _a ^8 = 8.0$~barn) and
the $^{12}$C was the major moderator (its microscopic scattering cross-section
was set at $\sigma _s ^{12} = 4.5$~barn). The $^{238}$U and $^{12}$C nuclei
concentrations ratio was set to the characteristic level for the dicarbides:
\begin{equation}
K_2 = \frac{N^{238}}{N^{12}} = 0.5 \nonumber
\end{equation}

The Fermi spectrum for the neutrons in moderating and absorbing medium of the
georeactor (carbon played a role of the moderator, and the $^{238}$U), $^{239}$U
and $^{239}$Pu played the role of the absorbers) was taken in the same 
form~(\ref{eq76}).

\begin{figure}[htb]
\centering
\includegraphics[width=12cm]{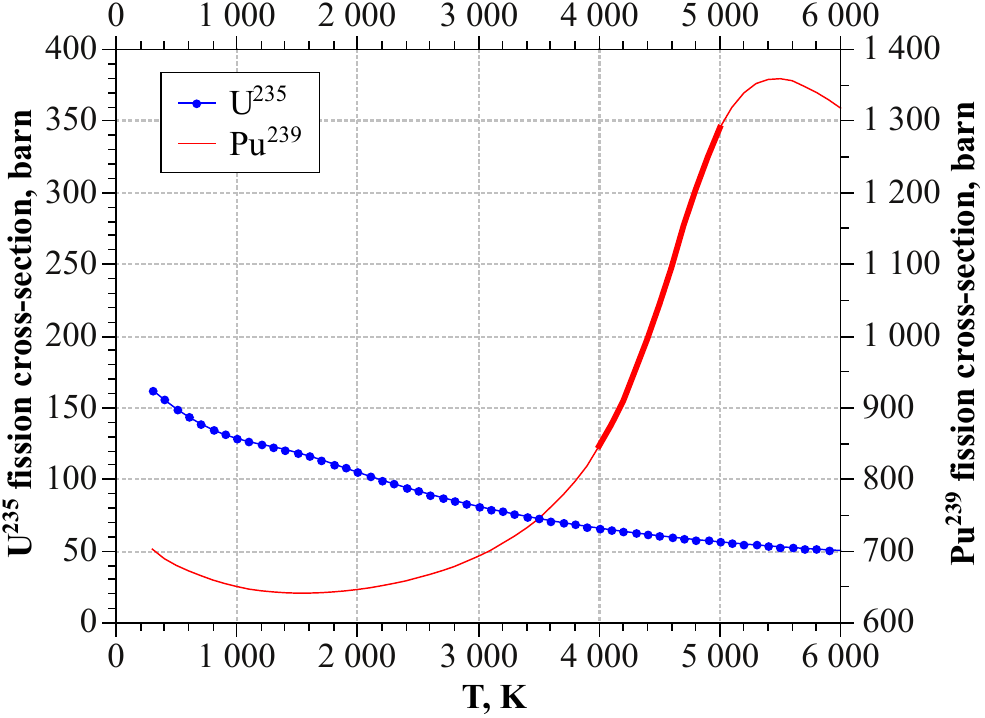}
\caption{The temperature dependences of the $^{239}$Pu fission cross-section 
         averaged over the neutron spectrum for the limit energy for the Fermi
	 and Maxwell spectra joining equal to 3kT. The analogous dependency for
	 the $^{235}$U is also shown.}
\label{fig17a}
\end{figure}

As en example Fig.~\ref{fig17a} shows the calculated temperature dependences
of the $^{235}$U and $^{239}$Pu fission cross-sections averaged over the neutron
spectrum.

The temperature choice is conditioned by the following important consideration: 
``Is it possible to obtain a solution (i.e. a spatio-temporal temperature 
distribution) in a form of the stationary solitary wave with a limited amplitude
instead of a $\delta$-function at some local spatial area, under such conditions
(6000K) emulating the time-conjugated heat takeoff (see fig.~\ref{fig12})?'' As 
shown below, such approach really works.

Below we present some calculation characteristics and parameters. During these 
calculations we used the following expression for dependence of the heat 
conductivity coefficient:

\begin{equation}
\aleph = 0.18 \cdot 10^{-4} \cdot T  \nonumber,
\end{equation}

\noindent which was obtained using the Wiedemann–Franz law and the data on 
electric conductivity of metals at temperature 6000K~\cite{ref71}. Specific heat
capacity at constant pressure was determined by value $c_p \approx 6~cal/(mol 
\cdot deg)$ according to Dulong and Petit law.

\begin{figure}
  \begin{center}
    \includegraphics[width=17cm]{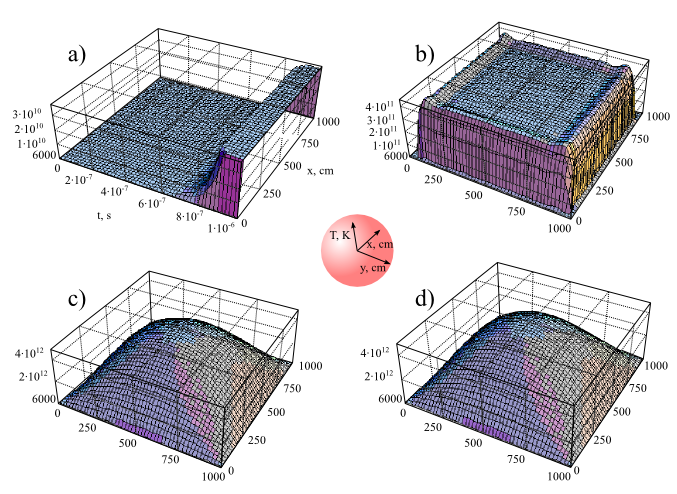}
    \caption{Heat transfer equation solution for a model georeactor (source 
             $\sim$ 2$^{nd}$ order temperature dependence, $const$ = 4.19~$J / 
	     (cm^3 \cdot s \cdot K^2)$; initial and boundary temperatures equal 
	     to 6000~K; fissile medium is a cube 10$\times$10$\times$10~m. The 
	     presented results correspond to the following times of temperature 
	     field evolution: (a) (1-10)$\cdot 10^{-7}$~s, (b) $10^{-6}$~s, 
	     (c) 0.5~s, (d) 50~s.}
    \label{fig18}
  \end{center}
\end{figure}

The fissile uranium-plutonium medium was modeled as a cube with dimensions 
10.0$\times$10.0$\times$10.0~m (fig.\ref{fig18}). Here for heat source we used 
the 2$^{nd}$ order temperature dependence (see (\ref{eq79})).
 
And finally fig.~\ref{fig18}a-d present a set of solutions of heat transfer 
equation (\ref{eq69}) with nonlinear exponential heat source (\ref{eq79}) in 
uranium-plutonium fissile medium with boundary and initial conditions emulating 
such process of heat takeoff that initial and boundary temperatures remain 
constant and equal to 6000K.

It is important to note here, that the solution set presented at fig.~
\ref{fig18}, demonstrates the solution tendency towards its ``stationary'' 
state quite clearly. This is achieved using the so-called ``magnifying glass''  
approach, when the solutions of the same problem are deliberately investigated 
at different timescales. E.g. fig.~\ref{fig18}a  shows the solution at the time 
scale $t \in [0,10^{-6}~s]$, while fig.~\ref{fig18}b describes the spatial 
solution of the problem (temperature field) for $t = 10^{-6}~s$. The fig.~
\ref{fig18}c-d presents the solution (spatial temperature distribution) at 
$t = 0.5~s$ and $t = 50~s$. 

As one can see, the solution (fig.~\ref{fig18}d) is completely identical to the 
previous (fig.\ref{fig18}c), i.e. to the distribution established in the medium 
in 0.5 seconds, which allowed us to make a conclusion on the temperature field 
stability, starting from some moment. It is interesting that the established 
temperature field creates the conditions suitable for the thermonuclear 
synthesis reaction, i.e. reaching 10$^{8}$K, and such temperature field lifetime
is not less than 50~s. These conditions are highly favorable for a stable 
thermonuclear burning, according to a known Lawson criterion, provided the
necessary nuclei concentration entering the thermonuclear synthesis reaction.

One should keep in mind though, that the results of this chapter are for the
purpose of demonstration only, since their accuracy is rather uncertain and 
requires a careful investigation with application of the necessary computational
resources. Nevertheless, the qualitative peculiarities of these solutions should
attract the researchers' attention to the nontrivial properties of the blow-up 
modes -- at least, with respect to the obvious problem of the inherent TWR 
safety violation.

\section{Conclusions}
\label{sec7}

Let us give some short conclusions stimulated by the following significant 
problems.

\begin{enumerate}
\item \textbf{TWR and the problem of dpa-parameter in cladding materials.} A 
possibility to surmount the so-called problem of dpa-parameter based on 
the conditions of nuclear burning wave existence in U-Pu and Th-U cycles is 
shown. In other words it is possible to find a nuclear burning wave mode, whose
parameters (fluence/neutron flux, width and speed of the wave) satisfy the 
dpa-condition (\ref{eq68a}) of the reactor materials radiation resistance, 
particularly, that of the cladding materials. It can be done using the joined 
application of the ``differential''\cite{ref29} and ``integral''\cite{ref20,
ref21,ref26} conditions for nuclear burning wave existence. The latter means 
that at the present time the problem of dpa-parameter in cladding materials in 
the TWR-project is not an insurmountable technical problem and can be 
satisfactorily solved.

Here we may add that this algorithm of an optimal nuclear burning wave mode 
selection predetermines a satisfactory solution of another technical problems 
mentioned in introduction. For example, the fuel rod length in the proposed TWR 
variant (see the ``ideal'' case in Table\ref{tab1}) is predetermined by the 
nuclear burning wave speed, which in a given case equals to 0.254~cm/day$\equiv$
85~cm/year, i.e. 20~years of TWR operation requires the fuel rod length $\sim$
17~m. On the other hand, it is known~\cite{ref72} that for a twisted fuel rod 
form with two- or four-bladed symmetry, the tension emerging from the fuel rod 
surface cooling is 30\% lower than that of a round rod with the same diameter, 
other conditions being equal. The same reduction effect applies to the hydraulic
resistance in comparison to a round rod of the same diameter.

Another problem associated with the reactor materials swelling is also solved 
rather simply. It is pertinent to note that if a ferritic-martensitic material 
is chosen as a cladding material (fig.~\ref{fig8}~\cite{ref48}), then the 
swelling effect at the end of operation will be only $\sim$0.5\%~\cite{ref48}. 
We could discuss other drawbacks mentioned in the introduction as well, but in 
our opinion, the rest of the problems are not the super-obstacles for the 
contemporary level of nuclear engineering, as compared to the main problem of 
dpa-condition, and can be solved in a traditional way.

\item \textbf{The consequences of the anomalous $^{238}U$ and $^{239}Pu$ 
cross-sections behavior with temperature.} It is shown that the capture and 
fission cross-sections of $^{238}U$ and $^{239}Pu$ manifest a monotonous growth 
in 1000-3000K range. Obviously, such anomalous temperature dependence of 
$^{238}U$ and $^{239}Pu$ cross-sections changes the neutron and heat kinetics of
the nuclear reactors drastically. It becomes essential to know their influence 
on kinetics of heat transfer because it may become the cause of a positive 
feedback with neutron kinetics, which may lead not only to undesirable loss of 
the nuclear burning wave stability, but also to a reactor runaway with a 
subsequent disaster.

\item \textbf{Blow-up modes and the problem of the nuclear burning wave 
stability.} One of the causes of possible fuel temperature growth is a 
deliberate or spontaneous coolant loss similar to Fukushima nuclear accident. As
shown above, the coolant loss may become a cause of the nonlinear heat source 
formation in the nuclear fuel and the corresponding mode with temperature and 
neutron flux blow-up. In our opinion, the preliminary results of heat transfer 
equation with nonlinear heat source investigations point to an extremely 
important phenomenon of the anomalous behaviour of the heat and neutron flux 
blow-up modes. This result poses a natural nontrivial problem of the fundamental
nuclear burning wave stability, and correspondingly, of a physically reasonable 
application of the Lyapunov method to this problem.

It is shown that some variants of the solution stability loss are caused by 
anomalous nuclear fuel temperature evolution. They can lead not only to the TWR 
inherent safety loss, but -- through a bifurcation of states (and this is very 
important!) -- to a new stable mode when the nuclear burning wave periodically 
``experiences'' the so-called ``controlled blow-up'' mode. At the same time, it 
is noted that such fast (blow-up regime) process has a number of physical 
uncertainties, which may happen to be experimentally insurmountable for the 
purposes of such process control.

\item \textbf{On-line remote neutrino diagnostics of the intra-reactor 
processes.} The high-power TWR or a nuclear fuel transmutation reactor are the 
projects with the single-load, fuel burn-up and the subsequent burial of the 
reactor apparatus. Thus an obvious necessity for the system of remote neutrino 
monitoring of the nuclear burning wave in the normal operation mode and the 
neutron kinetics in emergency situation. The details and peculiarities of the
isotope composition spatio-temporal distribution calculation in the active 
zone of the TWR are presented in~\cite{ref73,ref17,ref75} in detail within the 
inverse problem of the intra-reactor processes neutrino diagnostics.

\end{enumerate}

\bibliographystyle{elsarticle-num}
\bibliography{TravelingWaveReactor}

\end{document}